\documentclass[preprint,12pt]{elsarticle}
\usepackage{float,epsfig, floatflt,here}
\usepackage{rotating}
\usepackage{amssymb}
\usepackage{amsmath}

\begin{document}

\begin{frontmatter}
\title{A new family of implicit fourth order compact schemes for unsteady convection-diffusion equation with variable convection coefficient}
\author{Shuvam Sen}
\ead{shuvam@tezu.ernet.in}
\address{Department of Mathematical Sciences, Tezpur University, PIN 784028, INDIA}
\begin{abstract}
In this paper, a new family of implicit compact finite difference schemes for computation of unsteady convection-diffusion equation with variable convection coefficient is proposed. The schemes are fourth order accurate in space and second or lower order accurate in time depending on the choice of weighted time average parameter. The proposed schemes, where transport variable and its first derivatives are carried as the unknowns, combine virtues  of compact discretization and Pad\'{e} scheme for spatial derivative. These schemes which are based on five point stencil with constant coefficients, named as \emph{(5,5) Constant Coefficient 4th Order Compact} [(5,5)CC-4OC], give rise to a diagonally dominant system of equations and shows higher accuracy and better phase and amplitude error characteristics than some of the standard methods. These schemes are capable of using a grid aspect ratio other than unity and are unconditionally stable. They efficiently capture both transient and steady solutions of linear and nonlinear convection-diffusion equations with Dirichlet as well as Neumann boundary condition. The proposed schemes can be easily implemented and are applied to problems governed by incompressible Navier-Stokes equations apart from linear convection-diffusion equation. Results obtained are in excellent agreement with analytical and available numerical results in all cases, establishing efficiency and accuracy of the proposed scheme.
\end{abstract}
\begin{keyword}
Unsteady convection-diffusion equation \sep High order compact scheme \sep Navier-Stokes equations
\end{keyword}
\end{frontmatter}

\section{Introduction}
In this paper, we consider the unsteady two-dimensional (2D) convection-diffusion equation
\begin{equation}\label{1}
a\frac{\partial \phi}{\partial t}-\frac{\partial^2 \phi}{\partial x^2}-\frac{\partial^2 \phi}{\partial y^2}+c(x,y,t)\frac{\partial \phi}{\partial x}+d(x,y,t)\frac{\partial \phi}{\partial y}=s(x,y,t),\;\;\;\;\;(x,y,t)\in\Omega\times (0, T]
\end{equation}
for the unknown transport variable $\phi(x,y,t)$ in a rectangular domain $\Omega\subset\mathbb{R}^2$, $(0,T]$ is the time interval, with the initial condition
\begin{equation}\label{2}
\phi(x,y,0)=\phi_0(x,y),\;\;\;\;\;(x,y)\in\Omega
\end{equation}
and boundary condition
\begin{equation}\label{3}
b_1(x,y,t)\phi+b_2(x,y,t)\frac{\partial \phi}{\partial n}=g(x,y,t),\;\;\;\;\;(x,y)\in\partial\Omega,\;\;t\in(0,T].
\end{equation}
Here $a$ is a positive constant, $c(x,y,t)$ and $d(x,y,t)$ are convection coefficients, and $s(x,y,t)$ is a forcing function which together with $\phi_0(x,y)$ and $g(x,y,t)$ are assumed to be sufficiently smooth. $b_1$ and $b_2$ are arbitrary coefficients describing the boundary condition as a Dirichlet, Neumann, or Robin type in the boundary normal direction $n$.

Equation (\ref{1}) describes convection and diffusion of various physical properties such as mass, momentum, heat, and energy. It is encountered in many fields of science and engineering and often regarded as the simplified version of Navier-Stokes (N-S) equations. Numerical solution of the convection-diffusion equations plays an important role in computational fluid dynamics (CFD).

The emergence and growing popularity of compact schemes have brought about a renewed interest towards the finite difference (FD) approach. As such a great deal of effort towards numerical approximation of convection-diffusion equation using compact FD approach can now be seen in literature. A compact FD scheme is one which utilizes grid points located only directly adjacent to the node about which the differences are taken. These schemes offer higher accuracy even when the grid size is small. They are able to determine the flow with information solely from the nearest neighbours. The major advantage of compact discretization is that it leads to a system of equations with coefficient matrix having smaller band width as compared to non-compact schemes. For solving convection-diffusion equation, the high order compact (HOC) discretizations have been utilized in a number of different ways and a variety of specialized techniques have been developed. The pioneering works on unsteady 2D convection-diffusion equation were done by Hirsh \cite{hir_75} and Ciment \emph{et al.} \cite{cim_lev_wei_78}. In their works they have discussed conditionally stable compact FD schemes which are fourth order accurate in space and second order accurate in time. In 1984, Gupta \emph{et al.} \cite{gup_man_ste_84} proposed a compact FD scheme for the steady convection-diffusion equation with variable coefficients. Noye and Tan \cite{noy_tan_88} developed a spatially third order and temporally second order accurate nine point implicit HOC scheme with a large region of stability for 2D unsteady problem with constant coefficients. The last part of the previous century and early part of this century have seen significant development of compact algorithms for convection-diffusion equation in general and N-S equations in particular. In this regard the contributions of Dennis and Hudson \cite{den_hud_89}, Spotz and Carey \cite{spo_car_95,spo_car_01}, Li \emph{et al.} \cite{li_tan_for_95}, Zhang \cite{zha_98}, Kalita \emph{et al.} \cite{kal_dal_das_02}, Dehghan \cite{deh_04} deserves special mention. In the last decade or so HOC schemes for convection-diffusion equation have also been developed for nonuniform grids \cite{spo_car_98,kal_das_dal_04,wan_zho_zha_06}. Of late a lot of efforts have been made to increase computational efficiency of HOC schemes. This has led to the development and  implementation of HOC-ADI algorithms \cite{kar_zha_04,you_06,tia_ge_07,tia_11}. These HOC-ADI methods combine computational efficiency of ADI approach and high-order accuracy of HOC schemes. Besides, various multigrid and convergence acceleration techniques have been investigated and proposed to solve convection-diffusion equations discretized by the fourth order compact finite difference schemes \cite{gup_kou_zha_97,gup_zha_00,ge_cao_11}.

The aim of this paper is to introduce a new family of implicit compact FD formulation for solving unsteady convection-diffusion equation with variable convection coefficients. The developed schemes are fourth order accurate in space. One of the schemes also possess second order accuracy in time and rest are first order accurate. The approach involves discretizing the convection-diffusion equation using not only the nodal values of the unknown transport variable but also the values of its first derivatives at a few selected nodal points. Inturn first derivatives are discretized by using Pad\'{e} approximation. The new family of schemes are established on a rectangular domain using only the four nearest neighboring values of $\phi$. Although developed for convection-diffusion equation with variable coefficients the schemes are accessible in the form of a diagonally dominant system with sparse constant coefficient matrix and are named as \emph{(5,5) Constant Coefficient 4th Order Compact} [(5,5)CC-4OC] schemes. This significantly decreases computational complexity. In the process we also remove the usual restriction on HOC schemes of having to use a grid aspect ratio of unity. At first we develop a new HOC scheme for the 1D steady convection-diffusion equation to carry out wave number analysis and compare with other HOC schemes. Subsequently the family of implicit 4OC schemes are proposed and are shown to be unconditionally stable in von Neumann sense. To validate the accuracy and efficiency of the newly derived compact schemes numerical experiments are performed on five test problems. They are: (i) Taylor's vortex problem, (ii) The convection-diffusion of Gaussian pulse, (iii) Analytic solution of N-S equations with a source term, (iv) Doubly-periodic double shear layer problem, and (v) Lid driven cavity flow. Excellent comparison can be seen in all the cases.

The rest of this paper is organized into three sections. In section 2 we outline the (5,5)CC-4OC family of schemes and its properties. In section 3 we present numerical results and finally in section 4 concluding remarks are offered.

\section{The (5,5)CC-4OC schemes}
\subsection{Introduction of the new approach for steady 1D convection-diffusion equation}
In order to describe basic ideas of our approach we start with elementary 1D convection-diffusion equation
\begin{equation}\label{4}
-\phi_{xx}+c(x)\phi_{x}=s(x).
\end{equation}
A fourth order compact approximation of (\ref{4}) is obtained by considering the discretization
\begin{equation}\label{5}
\phi_{xx_i}=2\delta^2_x\phi_{_i}-\delta_x\phi_{x_{i}}+O(h^4)
\end{equation}
where $\delta_x$ and $\delta^2_x$ are the second order central difference operators for first and second order derivatives respectively, $h$ is the mesh length along $x$- direction and $\phi_i$ denote the approximate value of $\phi(x_i)$ at a typical grid point $x_i$. Using (\ref{5}) in (\ref{4}) and retaining the leading truncation errors we obtain a fourth order compact discretization
\begin{equation}\label{6}
-2\delta^2_x\phi_{i}+(c(x_i)+\delta_x)\phi_{x_{i}}=s_i+O(h^4).
\end{equation}
The main advantage of the discretized equation (\ref{6}) is that it leads to a diagonally dominant tridiagonal system with constant coefficient in transport variable $\phi$ and as such quite efficient for numerical computation. But the main disadvantage is that the derivative $\phi_{x_i}$ is required to be approximated upto fourth order of accuracy. This is separately achieved by using standard Pad\'{e} discretization
\begin{equation}\label{7}
\phi_{x_{i}}=(\delta_{x}\phi_{i}-\frac{h^2}{6}\delta^2_{x}\phi_{x_{i}})+O(h^4).
\end{equation}
The classical truncation error analysis shows that the relation (\ref{6}) along with (\ref{7}) constitute a compact fourth order discretization of the differential equation (\ref{4}). The idea of using derivative of flow variable in addition to the variable itself in the process of discretization, although new for convection-diffusion equation, has its roots in the work of Gupta and Manohar \cite{gup_man_79} on biharmonic equation. Later it was formulated by Stephenson \cite{ste_84}.

\subsection{Modified wave number analysis}\label{MWNA}
To examine in detail the characteristics of the fourth order compact scheme developed above, a modified wave number analysis of equation (\ref{4}) is carried out. Note that truncation error does not represent all the characteristics of a finite difference scheme and the modified wave number analysis provides a tool to compare the resolution of finite difference operator with its analytical counterpart. It allows one to assess how well different frequency components of a harmonic function in a periodic domain are represented by a discretization scheme \cite{you_06,tia_11}. Wave number analysis also provide an opportunity to compare this new philosophy \emph{vis-a-vis} with some of the well established finite difference algorithms.

Let us consider the trial function $e^{I\kappa x}$ ($I=\sqrt{-1}$) on a periodic domain where $\kappa$ is the wave number. The exact characteristic of the differential equation (\ref{4}) is
\begin{equation}\label{8}
\lambda_{Exact}=\kappa^2+Ic\kappa.
\end{equation}
We determine the characteristic of scheme (\ref{6}) by replacing the difference operators with the corresponding wave number and using the equation (\ref{7}). The characteristic of the newly proposed fourth order compact scheme is given by
\begin{equation}\label{9}
\lambda_{CC-4OC}=\frac{5-4\cos(\kappa h)-\cos^2(\kappa h)}{h^2(2+\cos(\kappa h))}+Ic\frac{3\sin(\kappa h)}{h(2+\cos(\kappa h))}.
\end{equation}

For the sake of comparison the characteristics of some of the well known schemes are presented below.

The characteristic of the second order central difference (CD) scheme
\begin{equation}\label{10}
(-\delta^2_x+c\delta_x)\phi_i=s_i
\end{equation}
for convection-diffusion equation (\ref{4}) is
\begin{equation}\label{11}
\lambda_{CD}=\lambda_1+Ic\lambda_2
\end{equation}
where $\displaystyle \;\;\;\lambda_1=\frac{2-2\cos(\kappa h)}{h^2}$, $\;\;\;\displaystyle \lambda_2=\frac{\sin(\kappa h)}{h}.\;\;\;$

The characteristic of fourth order HOC based schemes \cite{noy_tan_88,spo_car_01,kal_dal_das_02,kar_zha_04}
\begin{equation}\label{12}
(-\alpha_1\delta^2_x+c\delta_x)\phi_i=(1+\alpha_2\delta^2_x +\alpha_3\delta_x) s_i
\end{equation}
is
\begin{equation}\label{13}
\lambda_{HOC}=\frac{\alpha_1\lambda_1+Ic\lambda_2}{(1-\alpha_2\lambda_1)+I\alpha_3\lambda_2},
\end{equation}
where $\;\;\displaystyle\alpha_1=1+\frac{Pe^2}{12}$, $\;\;\;\;\;\;\;\;\;\;\;\;\displaystyle\alpha_2=\frac{1-\alpha_1}{c^2}+\frac{h^2}{6}$, $\;\;\;\;\;\;\;\;\;\;\;\;\;\;\;\displaystyle\alpha_3=\frac{1-\alpha_1}{c}$ and $Pe=ch$. $Pe$ is known as the cell Reynolds number.

The characteristic of the fourth order Pad\'{e} (PDE) approximation \cite{you_06}
\begin{eqnarray}\label{14}
-\phi_{xx_i}+c\phi_{x_i}&=&s_i\\
\phi_{xx_{i+1}}+10\phi_{xx_i}+\phi_{xx_{i-1}}&=&\frac{12}{h^2}(\phi_{_{i+1}}-2\phi_{_{i}}+\phi_{_{i-1}})\nonumber\\
\phi_{x_{i+1}}+4\phi_{x_i}+\phi_{x_{i-1}}&=&\frac{3}{h}(\phi_{_{i+1}}-\phi_{_{i-1}})\nonumber
\end{eqnarray}
is
\begin{equation}\label{15}
\lambda_{PDE}=\frac{12(1-\cos(\kappa h))}{h^2(2+\cos(\kappa h))}+Ic\frac{3\sin(\kappa h)}{h(2+\cos(\kappa h))}.
\end{equation}

Finally the characteristic of the RHOC scheme \cite{tia_11}
\begin{equation}\label{16}
(-\beta_1\delta^2_x+c\delta_x)\phi_i=(1+\beta_2\delta^2_x +\beta_3\delta_x) s_i
\end{equation}
is
\begin{equation}\label{17}
\lambda_{RHOC}=\frac{\beta_1\lambda_1+Ic\lambda_2}{(1-\beta_2\lambda_1)+I\beta_3\lambda_2},
\end{equation}
where
$\displaystyle \beta_1=\frac{1-\frac{Pe^2}{12}+\frac{Pe^4}{144}}{1-\frac{Pe^2}{6}+\frac{Pe^4}{36}}$,
$\displaystyle \;\;\beta_2=\Bigg\{%
\begin{array}{c}
  \frac{1-\beta_1}{c^2}+\frac{h^2}{6}\;\;\;c\neq0  \\
  \\
  \frac{h^2}{12}\;\;\;\;\;\;\;\;\;\;\;\;\;\;c=0  \\
\end{array}%
$,
$\displaystyle \;\;\beta_3=\Bigg\{%
\begin{array}{c}
  \frac{1-\beta_1}{c}\;\;\;c\neq0  \\
  \\
  0\;\;\;\;\;\;\;c=0.  \\
\end{array}%
$

\begin{figure}[!h]
\begin{minipage}[b]{.6\linewidth}\hspace{-1.cm}
\psfig{file=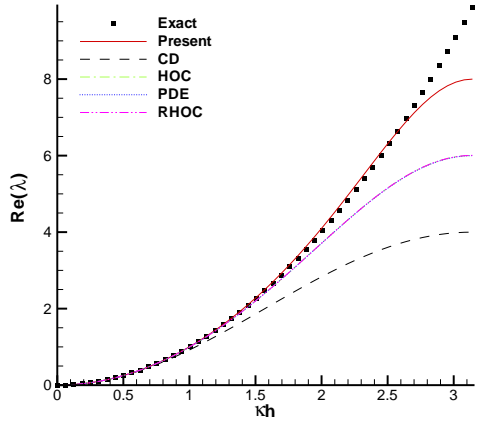,width=.9\linewidth}(a)
\end{minipage}
\begin{minipage}[b]{.6\linewidth}\hspace{-1.cm}
\psfig{file=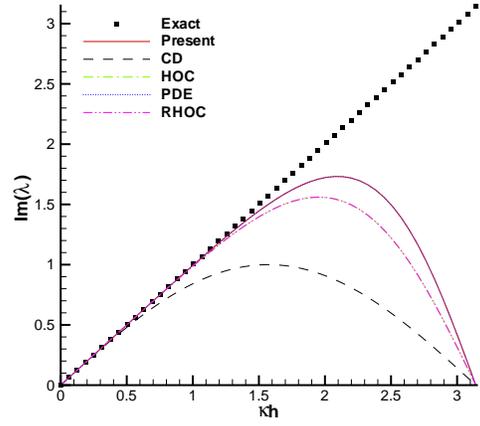,width=.9\linewidth}(b)
\end{minipage}
\begin{minipage}[b]{.6\linewidth}\hspace{-1.cm}
\psfig{file=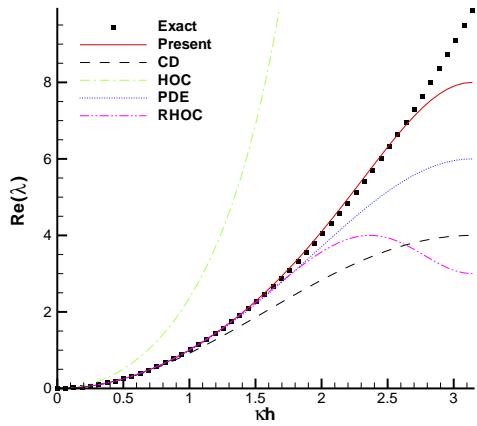,width=.9\linewidth}(c)
\end{minipage}
\begin{minipage}[b]{.6\linewidth}\hspace{-1.cm}
\psfig{file=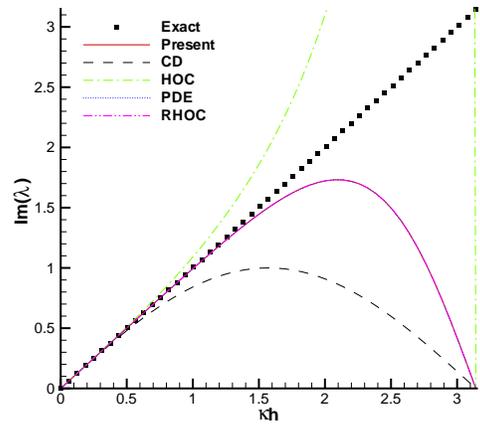,width=.9\linewidth}(d)
\end{minipage}
\begin{center}
\caption{{\sl The non-dimensional real and imaginary parts of $\lambda$ for five numerical schemes at two different cell Reynolds number $Pe=0.1$ ( (a), (b) ) and $Pe=100$ ( (c), (d) ).} }
\label{fig:char}
\end{center}
\end{figure}

Figure \ref{fig:char} shows the non-dimensionalised real and imaginary parts of characteristics ($\lambda$s) as function of $\kappa h$. We non-dimensionalise real and imaginary parts of $\lambda$s by multiplying with $h^2$ and $h/c$ respectively. Note that characteristics of the schemes (\ref{12}) and (\ref{16}) depend on $Pe$. Here we have considered two different cell Reynolds numbers $Pe=0.1$ and $Pe=100$. For the case $Pe=0.1$, the real part of the characteristic shown in figure \ref{fig:char}(a), the HOC, PDE and RHOC schemes are indistinguishable concluding identical dissipation errors whereas the second order accurate CD scheme shows much larger dissipation error. The figure clearly depicts that in comparison to all these schemes resolution of characteristic by the present compact scheme is much better and possess smaller dissipative error. The non-dimensional imaginary part of the characteristic corresponding to $Pe=0.1$ presented in figure \ref{fig:char}(b) shows that dispersive error of the present scheme is same as that of PDE scheme. This is also evident from equations (\ref{9}) and (\ref{15}). Both theses scheme show much smaller dispersive error when compared to HOC, RHOC and CD schemes. Note that at $Pe=0.1$ the imaginary part of the characteristic for HOC and RHOC schemes are basically same. As the cell Reynolds number is increased to $Pe=100$, the dissipation error for the HOC and RHOC scheme increases dramatically as shown in figure \ref{fig:char}(c), while the newly developed present scheme as also the PDE and the CD schemes do not alter their nondissipative properties. Again one can see the superiority of the present scheme. In figure \ref{fig:char}(d) we compare the dispersive errors of the schemes at $Pe=100$. One can clearly see a significant overshoot produced by HOC scheme. Here the present scheme produces the same resolution as that of the RHOC and the PDE schemes. The above discussion suggests that the newly developed compact scheme, which do not alter its characteristics with cell Reynolds number, possess resolution properties better than the schemes discussed here. Thus the present algorithm may be best suited for high Reynolds number flow computation.

\subsection{(5,5)CC-4OC scheme for the 2D unsteady case}
To begin with, we briefly discuss the development of HOC formulation for the steady state form of equation (\ref{1}), which is obtained when $c$, $d$, $s$ and $\phi$ are independent of $t$. Under these conditions, equation (\ref{1}) becomes
\begin{equation}\label{18}
-\frac{\partial^2 \phi}{\partial x^2}-\frac{\partial^2 \phi}{\partial y^2}+c(x,y)\frac{\partial \phi}{\partial x}+d(x,y)\frac{\partial \phi}{\partial y}=s(x,y).
\end{equation}
In order to obtain a compact spatially fourth order accurate discretization we consider the following approximations for second order space derivatives appearing in equation (\ref{18})
\begin{equation}\label{19}
\phi_{xx_{i,j}}=2\delta^2_x\phi_{_{i,j}}-\delta_x\phi_{x_{i,j}}+O(h^4),
\end{equation}
\begin{equation}\label{20}
\phi_{yy_{i,j}}=2\delta^2_y\phi_{_{i,j}}-\delta_y\phi_{y_{i,j}}+O(k^4).
\end{equation}
Here $h$ and $k$ are the mesh lengths along $x$- and $y$- directions respectively and $\phi_{i,j}$ denote the approximate value of $\phi(x_i, y_j)$ at a typical grid point $(x_i,y_j)$. The above approximations yield an $O(h^4, k^4)$ approximation for equation (\ref{18}) on a five point stencil as
\begin{equation}\label{21}
-2\delta^2_x\phi_{_{i,j}}-2\delta^2_y\phi_{_{i,j}}+(\delta_x+c_{i,j})\phi_{x_{i,j}}+(\delta_y+d_{i,j})\phi_{y_{i,j}}=s_{i,j}.
\end{equation}
Compatible fourth order approximations for space derivatives $\phi_{x_{i,j}}$ and $\phi_{y_{i,j}}$ appearing in equation (\ref{21}) may be obtained using
\begin{equation}\label{22}
\phi_{x_{i,j}}=(\delta_{x}\phi_{i,j}-\frac{h^2}{6}\delta^2_{x}\phi_{x_{i,j}}),
\end{equation}
and
\begin{equation}\label{23}
\phi_{y_{i,j}}=(\delta_{y}\phi_{i,j}-\frac{k^2}{6}\delta^2_{y}\phi_{y_{i,j}})
\end{equation}
respectively. Note that compared to standard HOC formulation \cite{kal_dal_das_02}, we are not required to assume any smoothness condition on the convection coefficients $c$, $d$ and forcing function $s$. The equation (\ref{21}) can be viewed as a banded system with only five non zero diagonals; of course drawback of requiring to approximate $\phi_{x_{i,j}}$ and $\phi_{y_{i,j}}$ separately using (\ref{22}) and (\ref{23}) respectively remain.

\begin{figure}[!h]
\begin{center}
\includegraphics[width=5in]{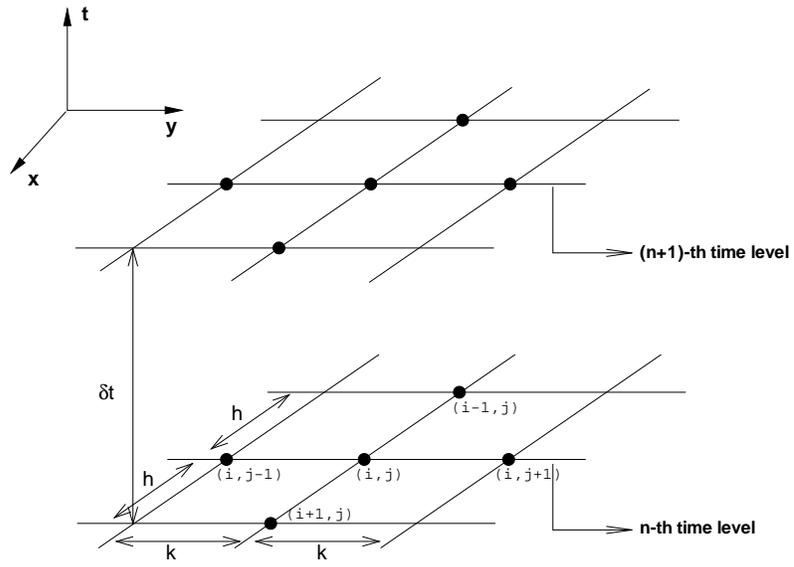}
\vspace{-1.5cm}
\caption{\sl Computational stencil for the scheme: the used nodes are denoted by ``$\bullet$".}
    \label{fig:stencil}
\end{center}
\end{figure}

For unsteady case, the equation with variable coefficients will be similar to equation (\ref{18}), but the coefficients $c$ and $d$ are functions of $x$, $y$ and $t$, and the expression on the right hand side becomes $\displaystyle s(x,y,t)-a\frac{\partial \phi}{\partial t}$. We intend to discretize time derivative as accurately as possible and obtain a stable numerical scheme. Introducing weighted time average parameter $\iota$ such that $t_{\iota} = (1 - \iota)t^{(n)}_{\iota} + \iota t^{(n+1)}_{\iota}$ for $0 \leqslant \iota \leqslant 1$, where $(n)$ denote the $n$-th time level, we obtain a family of integrators; for example, forward Euler for $\iota=0$, backward Euler for $\iota=1$ and Crank-Nicholson for $\iota=0.5$. Consequently the finite difference schemes for equation (\ref{1}) is
\begin{eqnarray}\label{24}
[a-2\iota\delta t(\delta^2_x+\delta^2_y)]\phi^{(n+1)}_{_{i,j}}&=&[a+2(1-\iota)\delta t(\delta^2_x+\delta^2_y)]\phi^{(n)}_{_{i,j}}\nonumber\\
 &+&(1-\iota)\delta t[-(\delta_x+c^{(n)}_{i,j})\phi^{(n)}_{x_{i,j}}-(\delta_y+d^{(n)}_{i,j})\phi^{(n)}_{y_{i,j}}+s^{(n)}_{i,j}]\nonumber\\
 &+&\iota\delta t[-(\delta_x+c^{(n+1)}_{i,j})\phi^{(n+1)}_{x_{i,j}}-(\delta_y+d^{(n+1)}_{i,j})\phi^{(n+1)}_{y_{i,j}}+s^{(n+1)}_{i,j}].\nonumber\\
\end{eqnarray}
The accuracy of the schemes are $O(\delta t^s, h^4, k^4)$, with $s\leqslant 2$. The second order accuracy in time is obtained for $\iota=0.5$. All the schemes arising in this way are implicit and in the following subsection it will be shown that this family of schemes is unconditionally stable for $0.5\leqslant\iota\leqslant 1$. For all values of $\iota$ in this range, except $\iota=1$, the difference stencil requires five points in both $n$-th and $(n+1)$-th time levels resulting in what we call $(5,5)$ scheme. The compact stencil emerging in this way has been illustrated in figure \ref{fig:stencil}.

\subsection{Stability analysis}
In this section we carry out von Neumann linear stability analysis of the finite difference scheme (\ref{24}) by assuming the convective coefficients $c$ and $d$ to be constants and forcing function $s$ in equation (\ref{1}) to be zero.
\newtheorem{thm}{Theorem}
\begin{thm}[Stability]\label{theorem1}
The finite difference scheme (\ref{24}) is unconditionally stable, in the von Neumann sense, for $0.5\leqslant\iota\leqslant 1$.
\end{thm}
\newproof{pf}{Proof}
\begin{pf}
Let $\phi^{(n)}_{i,j}=b^{(n)}e^{I\theta_{x}i}e^{I\theta_{y}j}$ where $b^{(n)}$ is the amplitude at time level $n$, and $\theta_{x}=2\pi h/\Lambda_1$, $\theta_{y}=2\pi k/\Lambda_2$ are the phase angles with wavelengths $\Lambda_1$ and $\Lambda_2$ respectively, then
from relations (\ref{22}) and (\ref{23}) we get
\begin{eqnarray}\label{25}
\phi_{{x}_{i,j}}^{(n)}=I\frac{3\sin\theta_{x}}{h(2+\cos\theta_{x})}\phi^{(n)}_{i,j}
\end{eqnarray}
and
\begin{eqnarray}\label{26}
\phi_{{y}_{i,j}}^{(n)}=I\frac{3\sin\theta_{y}}{k(2+\cos\theta_{y})}\phi^{(n)}_{i,j}.
\end{eqnarray}
Also
\begin{eqnarray}\label{27}
(\delta^2_x+\delta^2_y)\phi^{(n)}_{_{i,j}}=\bigg(\frac{2}{h^2}(\cos\theta_{x}-1)+\frac{2}{k^2}(\cos\theta_{y}-1)\bigg)\phi^{(n)}_{_{i,j}},
\end{eqnarray}
\begin{eqnarray}\label{28}
\delta_x\phi^{(n)}_{_{i,j}}=I\frac{\sin\theta_{x}}{h},
\end{eqnarray}
\begin{eqnarray}\label{29}
\delta_y\phi^{(n)}_{_{i,j}}=I\frac{\sin\theta_{y}}{k}.
\end{eqnarray}
Using relations (\ref{25})-(\ref{29}) in the scheme (\ref{24}) we obtain
\begin{eqnarray}
&&\bigg[a-4\iota\delta t\bigg(\frac{\cos\theta_{x}-1}{h^2}+\frac{\cos\theta_{y}-1}{k^2}\bigg)\bigg]\phi^{(n+1)}_{_{i,j}}\nonumber\\
&&=\bigg[a+4(1-\iota)\delta t\bigg(\frac{\cos\theta_{x}-1}{h^2}+\frac{\cos\theta_{y}-1}{k^2}\bigg)\bigg]\phi^{(n)}_{_{i,j}}\nonumber\\
&&+(1-\iota)\delta t\bigg[\bigg(\frac{\sin\theta_{x}}{h}-Ic\bigg)\frac{3\sin\theta_{x}}{h(2+\cos\theta_{x})}+\bigg(\frac{\sin\theta_{y}}{k}-Id\bigg)\frac{3\sin\theta_{y}}{k(2+\cos\theta_{y})}\bigg]\phi^{(n)}_{_{i,j}}\nonumber\\
&&+\iota\delta t\bigg[\bigg(\frac{\sin\theta_{x}}{h}-Ic\bigg)\frac{3\sin\theta_{x}}{h(2+\cos\theta_{x})}+\bigg(\frac{\sin\theta_{y}}{k}-Id\bigg)\frac{3\sin\theta_{y}}{k(2+\cos\theta_{y})}\bigg]\phi^{(n+1)}_{_{i,j}}\nonumber
\end{eqnarray}
If $G$ is the amplification factor then
\begin{eqnarray}
G&&=\frac{1+\frac{(1-\iota)\delta t}{a}
\bigg[\bigg(\frac{\cos^2\theta_{x}+4\cos\theta_{x}-5}{h^2(2+\cos\theta_{x})}+\frac{\cos^2\theta_{y}+4\cos\theta_{y}-5}{k^2(2+\cos\theta_{y})}\bigg)
-I\bigg(c\frac{3\sin\theta_{x}}{h(2+\cos\theta_{x})}+d\frac{3\sin\theta_{y}}{k(2+\cos\theta_{y})}\bigg)
\bigg]}
{1-\frac{\iota\delta t}{a} \bigg[\bigg(\frac{\cos^2\theta_{x}+4\cos\theta_{x}-5}{h^2(2+\cos\theta_{x})}+\frac{\cos^2\theta_{y}+4\cos\theta_{y}-5}{k^2(2+\cos\theta_{y})}\bigg)
-I\bigg(c\frac{3\sin\theta_{x}}{h(2+\cos\theta_{x})}+d\frac{3\sin\theta_{y}}{k(2+\cos\theta_{y})}\bigg)
\bigg]}\nonumber\\
&&=\frac{1+(1-\iota)(A+IB)}{1-\iota(A+IB)}\nonumber
\end{eqnarray}
where $$A=\frac{\delta t}{a} \bigg(\frac{\cos^2\theta_{x}+4\cos\theta_{x}-5}{h^2(2+\cos\theta_{x})}+\frac{\cos^2\theta_{y}+4\cos\theta_{y}-5}{k^2(2+\cos\theta_{y})}\bigg)$$
and $$B=\frac{\delta t}{a}\bigg(c\frac{3\sin\theta_{x}}{h(2+\cos\theta_{x})}+d\frac{3\sin\theta_{y}}{k(2+\cos\theta_{y})}\bigg).$$
Therefore $$|G|^2=\frac{(1+A-\iota A)^2+(1-\iota)^2 B^2}{(1-\iota A)^2+\iota^2 B^2}.$$
For stability we require $|G| \leqslant 1$ which gives
$$2A+(A^2+B^2)(1-2\iota)\leqslant 0.$$
Since $\displaystyle \frac{\cos^2\theta+4\cos\theta-5}{2+\cos\theta}\leqslant0$ $\;\;\forall\;\;$ $\theta\in[0,2\pi]$ we note that $A\leqslant 0$. Thus the above inequality is clearly satisfied if $$(1-2\iota)\leqslant 0\Rightarrow\iota\geqslant 0.5.$$
This completes the proof.
\end{pf}

\subsection{Solution of algebraic system}\label{SAS}
Let us now discuss the solution of algebraic systems associated with the finite difference approximation (\ref{24}). Introducing the notations $$
\Phi=(\phi_{1,1}, \phi_{1,2},...,\phi_{m,n})^T{\rm ,}\; \Phi_{x}=({\phi_{x}}_{1,1}, {\phi_{x}}_{1,2}, ..., {\phi_{x}}_{m,n})^T{\rm and}\;  \Phi_{y}=({\phi_{y}}_{1,1}, {\phi_{y}}_{1,2}, ..., {\phi_{y}}_{m,n})^T$$ the resulting system of equations in matrix form can be written as
\begin{eqnarray}\label{30}
M_1\Phi^{(n+1)}=F_1\big(\Phi^{(n)},\Phi^{(n)}_{x},\Phi^{(n)}_{y},\Phi^{(n+1)}_{x},\Phi^{(n+1)}_{y}\big).
\end{eqnarray}
For a grid of size $m \times n$, the matrix $M_1$ has the dimension $mn$. For the scheme developed here $M_1$ is a diagonally dominant constant banded matrix with five non zero diagonals. Also $\Phi^{(n)}$, $\Phi^{(n)}_{x}$, $\Phi^{(n)}_{y}$, $\Phi^{(n+1)}$, $\Phi^{(n+1)}_{x}$, $\Phi^{(n+1)}_{y}$ are all $mn$ component vectors. At any time step, once $\Phi^{(n)}$ has been approximated $\Phi^{(n)}_{x}$, $\Phi^{(n)}_{y}$ can be obtained by solving tridiagonal systems
\begin{eqnarray}\label{31}
M_2\Phi^{(n)}_{x}=F_2\big(\Phi^{(n)}\big)
\end{eqnarray}
\begin{eqnarray}\label{32}
M_3\Phi^{(n)}_{y}=F_3\big(\Phi^{(n)}\big)
\end{eqnarray}
respectively. The equations (\ref{31}) and (\ref{32}) are the corresponding matrix forms of the relations (\ref{22}) and (\ref{23}). System (\ref{31}) and (\ref{32}) being tridiagonal systems they can be solved using highly efficient numerical algorithms. Thus the main objective is to solve the system (\ref{30}), thereby evaluating unknown vector $\Phi^{(n+1)}$. Note that difficulty arises due the presence of $(n+1)$-th time level gradients of $\Phi$ on the right hand side of equation (\ref{30}) as those quantities will be available only after solving for transport variable at the $(n + 1)$-th time level. To overcome this difficulty we adopt correcting to convergence approach. The entire strategy can be summarised in the following algorithm:
\begin{enumerate}
\item Begin with $\Phi^{(n)}$.
\item Obtain $\Phi^{(n)}_{{x}}$ and $\Phi^{(n)}_{{y}}$ using (\ref{31}) and (\ref{32}) respectively.
\item Take $\Phi^{(n+1)}_{_{old}}=\Phi^{(n)}$, $\Phi^{(n+1)}_{{x}_{old}}=\Phi^{(n)}_{{x}}$, $\Phi^{(n+1)}_{{y}_{old}}=\Phi^{(n)}_{{y}}$.
\item Correct to $\Phi^{(n+1)}_{_{new}}$ using (\ref{30}).
\item Correct to $\Phi^{(n+1)}_{{x}_{new}}$, $\Phi^{(n+1)}_{{y}_{new}}$ using (\ref{31}) and (\ref{32}) respectively.
\item If $\|\Phi^{(n+1)}_{_{new}}-\Phi^{(n+1)}_{_{old}}\|<\epsilon$ then $\Phi^{(n+1)}=\Phi^{(n+1)}_{_{new}}$.
\item $\Phi^{(n+1)}_{_{old}}=\Phi^{(n+1)}_{_{new}}$, $\Phi^{(n+1)}_{{x}_{old}}=\Phi^{(n+1)}_{{x}_{new}}$, $\Phi^{(n+1)}_{{y}_{old}}=\Phi^{(n+1)}_{{y}_{new}}$ goto step 4.
\end{enumerate}
Direct solution of any of the above linear system is impractical because of huge size of the coefficient matrix. We solve the system using the biconjugate gradient stabilized (BiCGStab) \cite{book_kel_95} method without preconditioning, where, due to the compact nature of grid, it is easy to implement matrix vector multiplication $M_1\Phi$ without the need of storing any of the entries of the matrix $M_1$. The convergence criterion for BiCGStab iteration based on norm of residual depend on grid size and problem under consideration but a typical choice may be $10^{-10}$. Similarly, a typical stopping criterion for the inner iteration can be set at $\epsilon=10^{-12}$. We carry out our computations on a Pentium Dual-Core processor based PC with 4 GB RAM using double precision floating point arithmetic.

\section{Numerical Examples}
We conduct numerical tests to examine and compare the accuracy and efficiency of the present fourth order accurate scheme. We use the proposed scheme and obtain numerical results for five different problems of varying complexity. The first two of these five problems are concerned with convection-diffusion equation and the other three solves Navier-Stokes equations in two dimensions. For all the problems discussed here we consider uniform mesh ($h=k$) and take $\iota=0.5$ to ascertain second order temporal accuracy.

\subsection{Problem 1: Taylor's vortex problem}
\begin{figure}[!h]
\begin{minipage}[b]{.6\linewidth}\hspace{-1cm}
\psfig{file=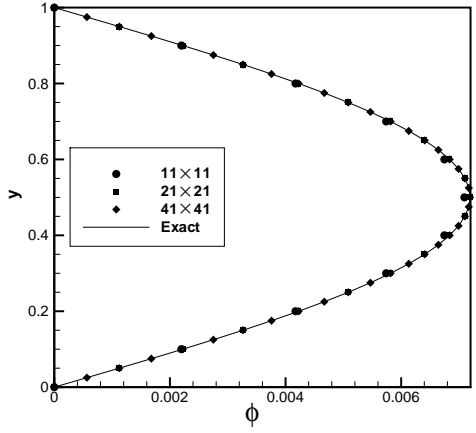,width=0.9\linewidth}
 \\(a)
\end{minipage}
\begin{minipage}[b]{.6\linewidth}\hspace{-1cm}
\psfig{file=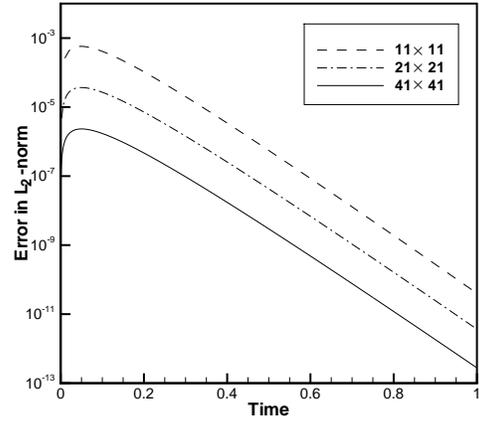,width=0.9\linewidth}
 \\(b)
\end{minipage}
\begin{center}
\caption{{\sl Problem 1: (a) Grid convergence along the line $x=0.5$ at $t=0.25$, (b) Time evolution of $L_2$- norm error for different grids.} }
    \label{fig:P1_grid_error}
\end{center}
\end{figure}

This is a pure convection problem in the unit square domain $[0,1]\times[0,1]$ with the coefficients $a=1$ and $c=d=0$. The exact solution of this test problem \cite{kar_zha_04,tia_ge_07,tia_11} is
\begin{equation}\label{33}
\phi(x,y,t)=e^{-2\pi^2t}\sin(\pi x)\sin(\pi y).
\end{equation}
In order to compute using discretized systems (\ref{24}) along with (\ref{22}) and (\ref{23}) the required initial and Dirichlet boundary conditions for $\phi$, $\phi_x$ and $\phi_y$ can be directly derived from (\ref{33}). We consider three different mesh sizes $11\times11$, $21\times21$ and $41\times41$ and compute errors with respect to the exact solution using $L_1$, $L_2$ and $L_{\infty}$ norms. The computed results at time $t=0.25$ and $t=0.5$ have been presented in table \ref{table:P1_order_spatial}. A spatial order of accuracy close to four, for all cases, can be seen in this table. We estimate order of accuracy using the formula $ln_2(Err1/Err2)$ where $Err1$ and $Err2$ are the errors with the grid sizes $h$ and $h/2$ respectively. Temporal second order accuracy of the scheme is verified in the table \ref{table:P1_order_temporal} where we compute using three different time steps $\delta t=0.01$, $\delta t=0.005$ and $\delta t=0.0025$ keeping $h$ unchanged. The grid convergence along the line $x=0.5$ have been presented in figure \ref{fig:P1_grid_error}(a) which establishes grid independence of the scheme. In figure \ref{fig:P1_grid_error}(b) we present time evolution of $L_2$- norm error for three different grid sizes from which stability and effectiveness of the scheme can be gauged.

\subsection{Problem 2: Convection-diffusion of Gaussian pulse}
\begin{figure}[!ht]
\begin{center}
\includegraphics[width=4.0in]{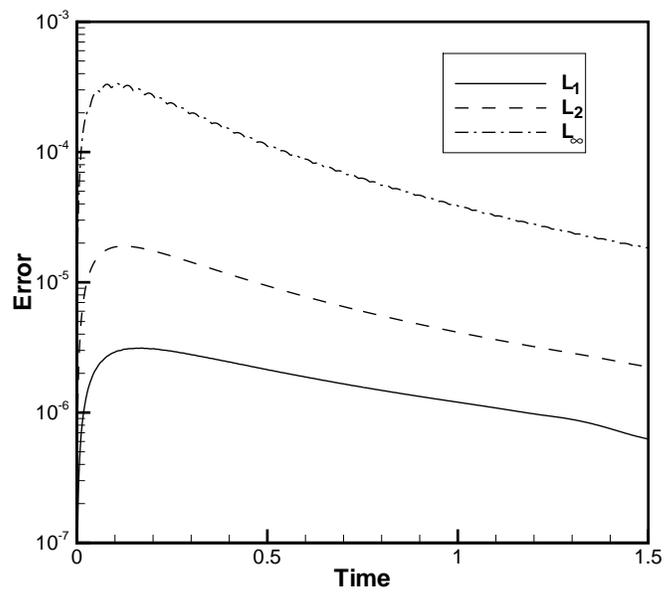}
\caption{\sl Problem 2: Time evolution of errors corresponding to three different norms. $Pe=2$, $a=100$, $c=d=80$, $\delta t=h^2=k^2$.}
    \label{fig:P2_error}
\end{center}
\end{figure}

To further study the effectiveness and validity of the new scheme we apply it to an unsteady problem concerning the convection-diffusion of a Gaussian pulse \cite{kar_zha_04,you_06,tia_11} in the square region $[0,2]\times[0,2]$ with analytical solution
\begin{equation}\label{34}
\phi(x,y,t)=\frac{1}{4t+1}exp\bigg[-\frac{(ax-ct-0.5a)^2}{a(4t+1)}-\frac{(ay-dt-0.5a)^2}{a(4t+1)} \bigg].
\end{equation}
The initial condition $\phi(x,y,0)$, taken from (\ref{34}), is a Gaussian pulse centered at ($0.5, 0.5$) with pulse height 1. Dirichlet boundary conditions are also directly obtained from the analytical solution. In this study we consider two different combinations of convection coefficients $c=d=80$ and $c=d=10000$ with fixed value of $a=100$. For the first case we compute using three different grids and present errors calculated using three different norms at time $t=0.5$ and $t=1.0$ in the table \ref{table:P2_order_spatial}. This table again verifies the analytical fourth order spatial convergence of the scheme. Using identical combination of the coefficients we estimate the temporal order of accuracy of the scheme in table \ref{table:P2_order_temporal}. In this table we have used three different time steps $\delta t=0.1$, $\delta t=0.05$, $\delta t=0.025$ keeping spatial grid spacing fixed at $h=k=0.05$. An order of accuracy around two can be seen in all the cases. The temporal decaying nature of error computed using three different norms for cell Reynolds number $Pe=2$ has been shown in figure \ref{fig:P2_error}. As expected all the norms show similar type of decaying character with time.

\begin{figure}[!h]
\begin{minipage}[b]{.6\linewidth}\hspace{-1cm}
\psfig{file=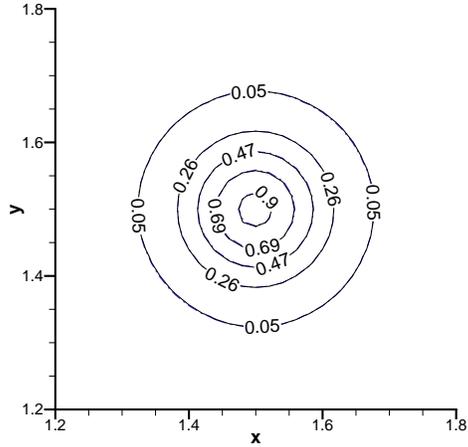,width=0.95\linewidth}
 \\(a)
\end{minipage}
\begin{minipage}[b]{.54\linewidth}\hspace{-1cm}
\psfig{file=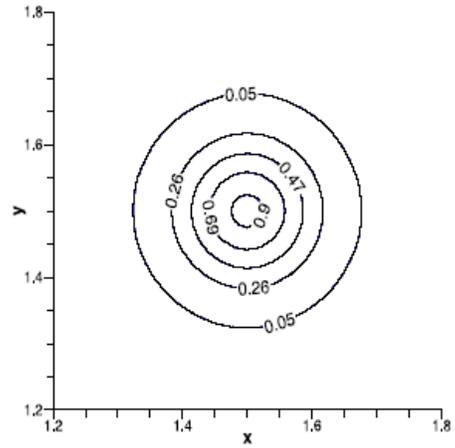,width=0.9\linewidth}
 \\(b)
\end{minipage}
\begin{minipage}[b]{.54\linewidth}\hspace{-1cm}
\psfig{file=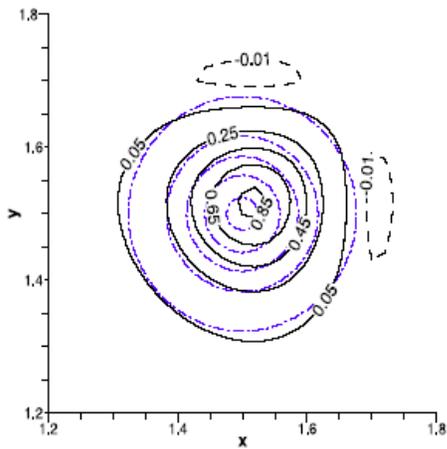,width=0.9\linewidth}
 \\(c)
\end{minipage}
\begin{minipage}[b]{.54\linewidth}
\psfig{file=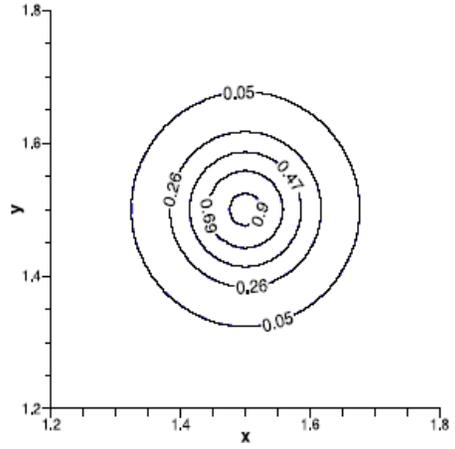,width=0.9\linewidth}
 \\(d)
\end{minipage}
\begin{center}
\caption{{\sl Problem 2: (a) Present, (b) PDE \cite{you_06}, (c) HOC \cite{kar_zha_04}, (d) RHOC \cite{tia_11}. $Pe=200$, $a=100$, $c=d=10^4$, $\delta t=2.5\times 10^{-5}$, $h=k=0.02$. Dash-dot contour lines in (a)-(d) correspond to the exact solution.} }
    \label{fig:P2_contour_com}
\end{center}
\end{figure}

The second combination of coefficients \emph{viz.} $a=100$, $c=d=10000$ has been motivated by the works of \cite{kar_zha_04,you_06,tia_11}. In figure \ref{fig:P2_contour_com}(a) we compare the analytical solution and the computed solution in the region $1.2\leqslant x$, $y\leqslant1.8$. This figure clearly shows that our computed solution is almost indistinguishable from the exact one. For this computation we have kept cell Reynolds number fixed at $Pe=200$ and have used $\delta t=2.5\times10^{-5}$. Figures \ref{fig:P2_contour_com}(b)-(d) obtained from \cite{tia_11} present similar comparisons for the schemes given by You \cite{you_06}, Karaa and Zhang \cite{kar_zha_04} and Tian \cite{tia_11} respectively. It is seen that although schemes given by You \cite{you_06} and Tian \cite{tia_11} captures the analytical solution accurately but the scheme of Karaa and Zhang \cite{kar_zha_04} fails to obtain the desired accuracy mainly due to the enhanced numerical dissipation. Note that in section \ref{MWNA} we have established enhanced phase error and hence numerical dissipation associated with the HOC type schemes discussed in \cite{kar_zha_04}.

\subsection{Problem 3: Analytic solution of N-S equations with a source term}
Next we consider the Navier-Stokes system of equations with a source term given by
\begin{eqnarray}\label{35}
\frac{\partial \omega}{\partial t}+u\omega_x+v\omega_y-\frac{1}{Re}(\omega_{xx}+\omega_{yy})&=&\frac{16}{Re}(x^2+y^2+4)e^{-\frac{t}{Re}}
\end{eqnarray}
\begin{eqnarray}\label{36}
-(\psi_{xx}+\psi_{yy})&=&\omega.
\end{eqnarray}
This problem considered in \cite{ben_cro_fis_05,kal_gup_10}, is defined on the square domain $[0, 1]\times[0, 1]$ and has analytical solution $\displaystyle \psi=(x^2+y^2)^2e^{-\frac{t}{Re}}$, $\omega=-16(x^2+y^2)e^{-\frac{t}{Re}}$, which decays rapidly with time. The presence of $\displaystyle u=\psi_y$ and $v=-\psi_x$ introduces non-linearity into the system. This problem is studied to discusses applicability and effectiveness of the newly developed formulation in tackling coupled non-linear N-S system. We begin by observing that formulation (\ref{24}) can be directly applied to equation (\ref{35}) where $\omega$ plays the role of transport variable. On the other hand for equation (\ref{36}) we use steady formulation (\ref{21}) with $c_{i,j}=d_{i,j}=0$ $\forall\; i, j$. The necessity to compute $\psi_x$ and $\psi_y$ using Pad\'{e} approximations (\ref{22}) and (\ref{23}) alleviates the need to compute $u$ and $v$ separately although one still has to do some additional computations for $\omega_x$ and $\omega_y$. At each time step we solve all systems concurrently in tandem, using the strategy outlined in section \ref{SAS}, where we impose convergence criterion on $\psi$. Initial and boundary conditions necessary for computation are derived from analytical expressions. Table \ref{table:P3_order_spatial_st} and \ref{table:P3_order_spatial_om} display $\psi$ and $\omega$ errors respectively as measured with respect to three different norms. The computations have been carried out by taking $Re=1$. We present results for three different time levels and see that the convergence rate is around four for both the $\psi$ and $\omega$. From the data presented in table \ref{table:P3_order_spatial_st} and \ref{table:P3_order_spatial_om} it is noted that our $L_{\infty}$ norm errors for $\psi$ and $\omega$ are much better than the errors reported in \cite{kal_gup_10}. At time $t=1.0$, our $\psi$ and $\omega$ errors are respectively about one-tenth and one-hundredth of the errors reported in \cite{kal_gup_10}.

\subsection{Problem 4: Doubly-periodic double shear layer problem}
The doubly periodic double shear layer flow problem has been used by authors \cite{bel_col_89,tia_lia_yu_11} to test the behavior of discretization for non-stationary situations where steep gradients are involved. The shear layers are perturbed slightly at the initial time, which causes it to roll up in time to large vortical structures. The unsteady incompressible N-S equations
\begin{eqnarray}\label{37}
\frac{\partial \omega}{\partial t}+u\omega_x+v\omega_y-\frac{1}{Re}(\omega_{xx}+\omega_{yy})&=&0
\end{eqnarray}
\begin{eqnarray}\label{38}
-(\psi_{xx}+\psi_{yy})&=&\omega.
\end{eqnarray}
are solved to obtain flow field. The initial conditions for $u$ and $v$ are given by
\begin{eqnarray}\label{39}
u(x,y,0)&=&
  \begin{cases}
   \tanh[(y-\pi/2)/\rho]   & \text{if } 0\leqslant y\leqslant \pi \\
   \tanh[(3\pi/2-y)/\rho] & \text{if } \pi\leqslant y\leqslant 2\pi
  \end{cases}\\
v(x,y,0)&=&\sigma \sin(x)\nonumber
\end{eqnarray}
from which expressions for $\psi$ and $\omega$ are obtained. We consider shear layer width parameter $\rho=\pi/15$ and size of perturbation $\sigma=0.05$ along with $Re=10000$ which is an excellent case to verify the precision and stability of the
numerical scheme at high wavenumbers. Periodic boundary conditions are imposed along both $x$ and $y$ directions.

\begin{figure}[!h]
\begin{minipage}[b]{.8\linewidth}\hspace{-1cm}
\centering\psfig{file=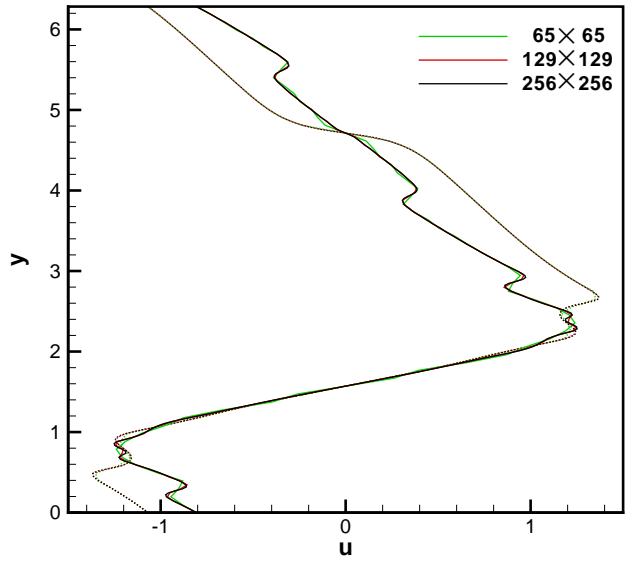,width=0.9\linewidth}
 \\(a)
\end{minipage}
\begin{minipage}[b]{.8\linewidth}\hspace{-1cm}
\centering\psfig{file=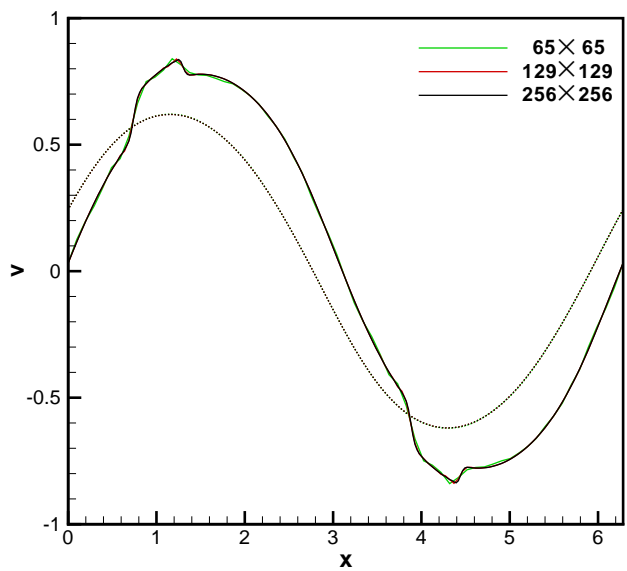,width=0.9\linewidth}
 \\(b)
\end{minipage}
\begin{center}
\caption{{\sl Problem 4: Gird convergence for velocity components at time $t=6.0$ (denoted by dotted lines) and $t=10.0$ (denoted by solid lines) (a) horizontal velocity along the vertical centreline and (b) vertical velocity along the horizontal centreline.} }
    \label{fig:P4_uv_com}
\end{center}
\end{figure}

\begin{figure}[!h]
\begin{minipage}[b]{.6\linewidth}
\centering\psfig{file=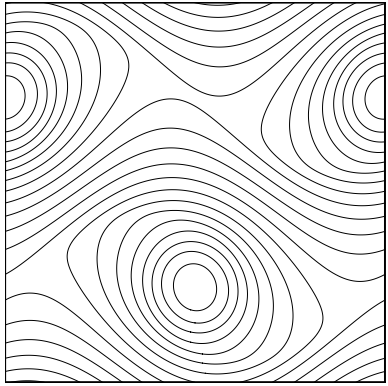,width=0.9\linewidth}
 \\(a)
\end{minipage}            \hspace{-2.5mm}
\begin{minipage}[b]{.6\linewidth}
\centering\psfig{file=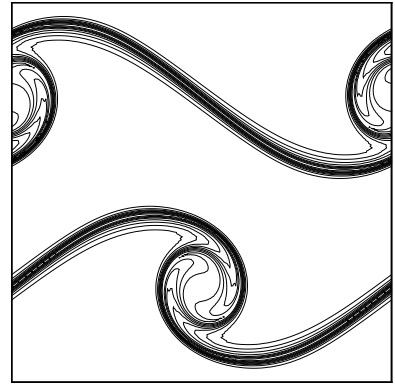,width=0.9\linewidth}
 \\(b)
\end{minipage}            \hspace{-2.5mm}
\begin{minipage}[b]{.6\linewidth}   \
\centering\psfig{file=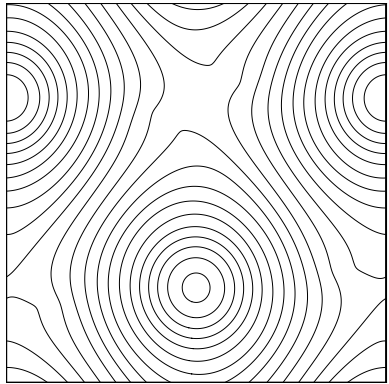,width=0.9\linewidth}
 \\(c)
\end{minipage}            \hspace{-2.5mm}
\begin{minipage}[b]{.6\linewidth}
\centering\psfig{file=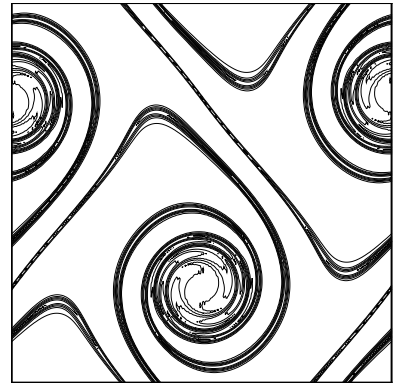,width=0.9\linewidth}
 \\(d)
\end{minipage}            \hspace{-2.5mm}
\begin{center}
\caption{{\sl Problem 4: (a) Streamline contours at time $t=6.0$, (b) Vorticity contours at time $t=6.0$, (c) Streamline contours at time $t=10.0$, (d) Vorticity contours at time $t=10.0$.} }
\label{fig:P4_stream_vort}
\end{center}
\end{figure}

The calculated values of the horizontal velocity $u$ along the vertical centerline and the vertical velocity $v$ along the horizontal centerline at time $t =6.0$ and $t=10.0$ are presented in figures \ref{fig:P4_uv_com}(a) and \ref{fig:P4_uv_com}(b) for the grids $65\times65$, $129\times129$, and $257\times257$. These figures verify the grid independence of the numerical results obtained using this newly developed algorithm. We see that all the sharp velocity gradients have been captured accurately by the grid of size $129\times129$ and is enough for accurate resolution of the flow. For this problem, in the absence of analytical solutions, we estimate the perceived order of convergence $p$ by using the formula \cite{book_roa_98}
\begin{equation}\label{40}
\frac{\|\Psi_3-\Psi_1\|}{\|\Psi_3-\Psi_2\|}\doteq\frac{h_1^{p}-h_3^{p}}{h_2^{p}-h_3^{p}}.
\end{equation}
Here $h_1$, $h_2$, $h_3$ are mesh sizes in three different uniform grids used to compute solutions vectors $\Psi_1$, $\Psi_2$, $\Psi_3$ respectively. In table \ref{table:P4_u_order} we present differences in the horizontal velocity fields computed using three different grids $65\times65$, $129\times129$, and $257\times257$. We calculate differences using $L_1$ and $L_2$ norms at time $t=1.0$, $t=5.0$, and $t=10.0$ and estimate perceived order of convergence by using (\ref{40}). An identical estimation has also been carried out for the vertical velocity field in table \ref{table:P4_v_order}. From these tables we see that at time $t=1.0$ the perceived order of convergence compares well with the analytical order of convergence, but with time progressing there is a drop in the perceived order of convergence for both $u$ and $v$ which may be attributed to the increasing complexity of the flow field due to severe roll up of the shear layer. Finally stream function and vorticity fields at time $t=6.0$ and $t=10.0$ computed using $257\times257$ grid is presented in figure \ref{fig:P4_stream_vort}. It is seen that a very good resolution is obtained by using our scheme. The present results are smooth and the integration is stable which is also sustained by the figure \ref{fig:P4_uv_com}. These results show that the present method is efficient for solving problems with steep gradients.

\subsection{Problem 5: Lid-driven cavity flow}
We turn now to the problem of 2D lid-driven cavity, which has been used as benchmark test problem by many authors. This problem is of great scientific interest because it displays almost all of the fluid mechanical phenomena for incompressible viscous flows in the simplest of geometric
settings. Over the years, this problem has become one of the most frequently used benchmark problem for the assessment of numerical methods, particularly for the steady state solution of the incompressible fluid flows governed by the N-S equations (\ref{37})-(\ref{38}). In particular, we compare our results to the steady state results of Ghia \emph{et al.} \cite{ghi_ghi_shi_82}, Bruneau and Jouron \cite{bru_jou_90}, and Gupta and Kalita \cite{gup_kal_05}. The cavity is defined as the square $0\leqslant x$, $y\leqslant1$. The fluid motion is generated by the sliding motion of the top wall of the cavity in its own plane from left to right. Boundary conditions on the top wall ($y =1$) are given as $u=1$, $v=0$. All other walls of the cavity are at rest.

We obtain the steady state solutions using a time-marching strategy. The steady state was assumed to reach when the maximum $\psi$-error between two successive outer temporal iteration steps was smaller than $1.0\times10^{-10}$. Numerical solutions for the driven-cavity flow are obtained taking Reynolds number $Re=1000$, $3200$, $5000$, and $7500$. The Reynolds number for the lid-driven cavity problem is defined as $Re=UL/\nu$, where $U=1$ is the characteristic velocity, $L=1$ is the characteristic length and $\nu$ is the kinematic viscosity. In the present computation we consider an uniform grid spacing with mesh size $\displaystyle h=k=\frac{1}{64}$ for $Re=1000$ and $\displaystyle h=k=\frac{1}{128}$ for the other three cases. The time increment for all the cases is equal to $\delta t=0.01$.

\begin{figure}[!h]
\begin{minipage}[b]{.8\linewidth}\hspace{-1cm}
\centering\psfig{file=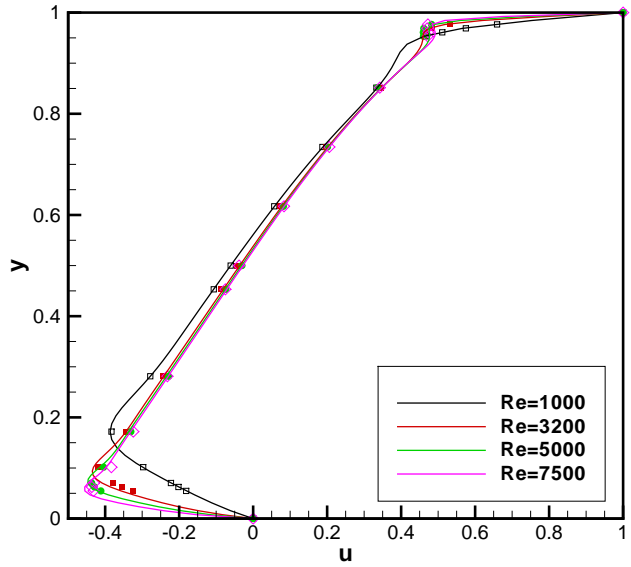,width=0.9\linewidth}
 \\(a)
\end{minipage}
\begin{minipage}[b]{.8\linewidth}\hspace{-1cm}
\centering\psfig{file=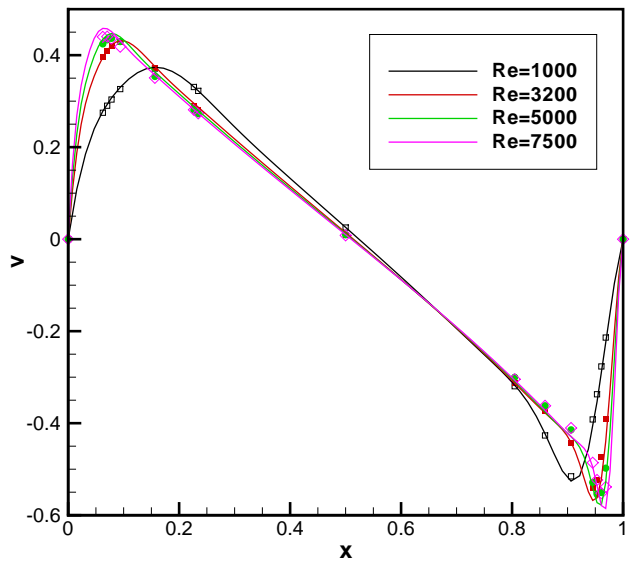,width=0.9\linewidth}
 \\(b)
\end{minipage}
\begin{center}
\caption{{\sl Problem 5: Comparisons of computed profiles (denoted by solid lines) of (a) horizontal velocity along the vertical centreline and (b) vertical velocity along the horizontal centreline with those of Ghia \emph{et al.} \cite{ghi_ghi_shi_82} (denoted using symbols) for different Reynolds number.} }
    \label{fig:P5_uv_com}
\end{center}
\end{figure}

In figure \ref{fig:P5_uv_com}, we present comparisons of our steady state results of the horizontal velocities on the vertical centerline and the vertical velocities on the horizontal centerline of the square cavity for Reynolds numbers $1000$ to $7500$ with the data from Ghia \emph{et al.} \cite{ghi_ghi_shi_82}. In each case, velocity profiles obtained by the proposed method on relatively coarser grids match very well with results given by Ghia \emph{et al.} Note that for $Re=1000$, $5000$ and $7500$ we have used mesh spacing twice that of Ghia \emph{et al.}

\begin{figure}[!h]
\begin{minipage}[b]{.6\linewidth}
\centering\psfig{file=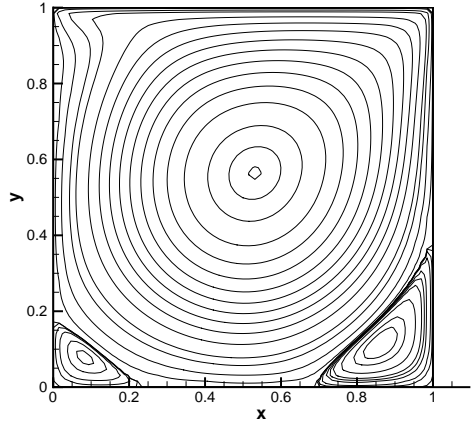,width=0.9\linewidth}
 \\(a)
\end{minipage}            \hspace{-2.5mm}
\begin{minipage}[b]{.6\linewidth}
\centering\psfig{file=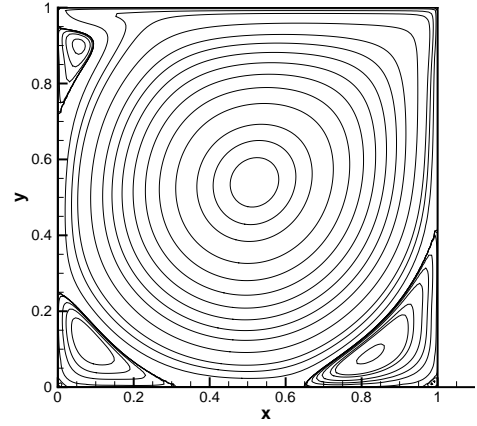,width=0.9\linewidth}
 \\(b)
\end{minipage}            \hspace{-2.5mm}
\begin{minipage}[b]{.6\linewidth}   \
\centering\psfig{file=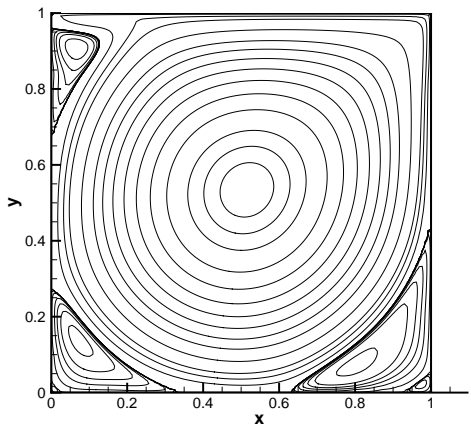,width=0.9\linewidth}
 \\(c)
\end{minipage}            \hspace{-2.5mm}
\begin{minipage}[b]{.6\linewidth}
\centering\psfig{file=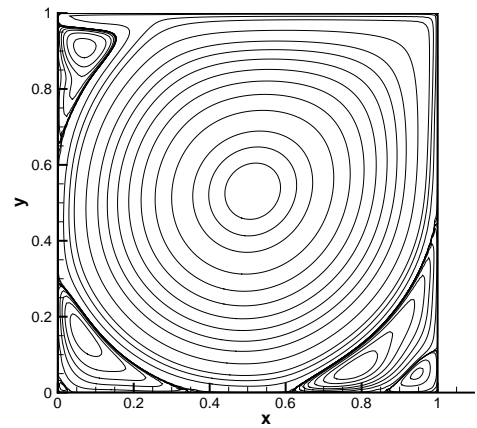,width=0.9\linewidth}
 \\(d)
\end{minipage}            \hspace{-2.5mm}
\begin{center}
\caption{{\sl Problem 5: The steady state streamline contours: (a) $Re=1000$, (b) $Re=3200$, (c)
$Re=5000$, and (d) $Re=7500$.} }
\label{fig:P5_stream}
\end{center}
\end{figure}

\begin{figure}[!h]
\begin{minipage}[b]{.6\linewidth}
\centering\psfig{file=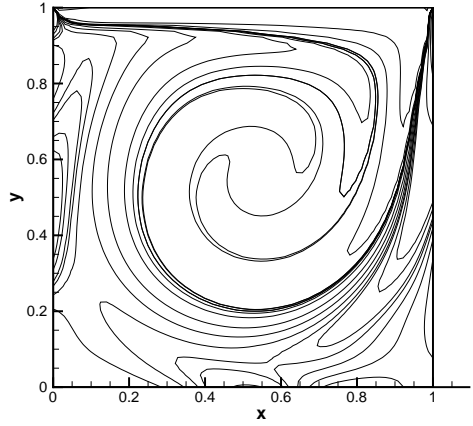,width=0.9\linewidth}
 \\(a)
\end{minipage}            \hspace{-2.5mm}
\begin{minipage}[b]{.6\linewidth}
\centering\psfig{file=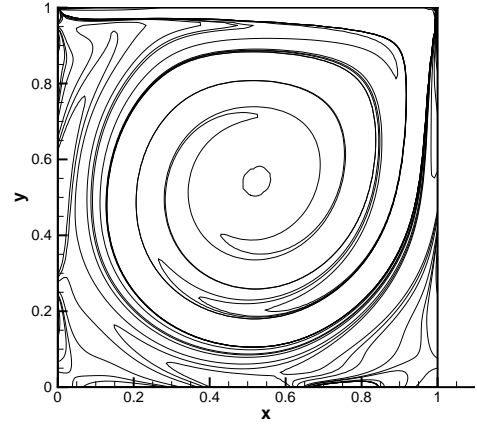,width=0.9\linewidth}
 \\(b)
\end{minipage}            \hspace{-2.5mm}
\begin{minipage}[b]{.6\linewidth}   \
\centering\psfig{file=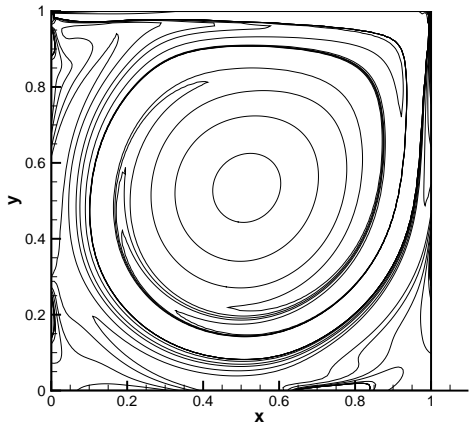,width=0.9\linewidth}
 \\(c)
\end{minipage}            \hspace{-2.5mm}
\begin{minipage}[b]{.6\linewidth}
\centering\psfig{file=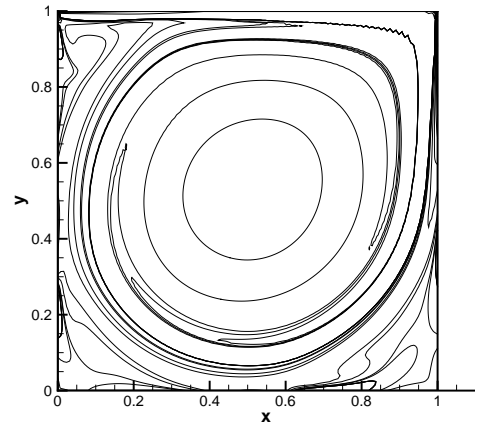,width=0.9\linewidth}
 \\(d)
\end{minipage}            \hspace{-2.5mm}
\begin{center}
\caption{{\sl Problem 5: The steady state vorticity contours: (a) $Re=1000$, (b) $Re=3200$, (c)
$Re=5000$, and (d) $Re=7500$.} }
\label{fig:P5_vorti}
\end{center}
\end{figure}

Figure \ref{fig:P5_stream} depicts streamline contours for Reynolds numbers $1000$, $3200$, $5000$, and $7500$. In these graphs, the typical separations and secondary vortices at the bottom corners of the cavity as well as at the top left can be seen. For $Re=7500$ a tertiary vortex can also be seen at the bottom right corner. The vorticity contours corresponding to the above mentioned $Re$ values can be found in figure \ref{fig:P5_vorti}. In this figure we have clearly depicted vorticity contours in the vicinity of the solid walls as well as at the centre of the cavity. The stream lines and vorticity contours obtained here are in good agreement with the established results available in literature \cite{tia_lia_yu_11,ghi_ghi_shi_82,bru_jou_90,gup_kal_05}, thereby confirming that our formulation yields qualitatively accurate solutions.

To further validate the present method we provide quantitative data comparison of our solutions in table \ref{table:P5_com}. Here we present our steady state data for the primary, secondary and tertiary vortex centres and compare them with the well established work of Ghia \emph{et al.} \cite{ghi_ghi_shi_82}, Bruneau and Jouron \cite{bru_jou_90}, and Gupta and Kalita \cite{gup_kal_05}. A very close comparison can be seen except possibly for the strength of the tertiary vortices which may be attributed to the coarser nature of the grid being used. It is clear from the table that the results of the present numerical method are reliable and the algorithm can be used to solve unsteady viscous incompressible flows.

\section{Conclusion}

A new family of implicit spatially fourth order accurate compact finite difference schemes have been proposed with weighted time discretization for unsteady convection-diffusion equation with variable convection coefficients. The schemes are second or lower order accurate in time according as time discretization parameter $\iota=0.5$ or otherwise. Linear stability analysis shows that for $0.5\leqslant\iota\leqslant1$ the schemes are unconditionally stable. The novelty of the family of schemes lies in the fact that the second order spatial derivatives of the transport variable are discretized using not only the nodal values of the transport variable but also its first derivatives leading to fourth order approximation. Pad\'{e} approximation used only for first derivatives ensures overall fourth order accuracy and diagonal dominance of the resultant constant coefficients algebraic systems. Modified wave number analysis confirms that the family of schemes enjoys better resolution properties and hence lesser dissipation error when compared to other compact schemes. The newly developed family of schemes have been used to solve Navier-Stokes system also and show excellent performance in terms of computational accuracy, efficiency, and stability with smaller number of grid points. Five different numerical experiments performed to demonstrate high accuracy and efficiency of the present method produce excellent comparison. The extension of the proposed discretization procedure to irregular domain and three dimensional cases are straightforward.

\bibliographystyle{plain}

\begin{thebibliography}{}
\bibitem{hir_75}
R.S. Hirsh, Higher order accurate difference solutions of fluid mechanics problems
by a compact differencing technique, Journal of Computational Physics 19 (1975) 90-109.

\bibitem{cim_lev_wei_78}
M. Ciment, S.H. Leventhal, B.C. Weinberg, The operator compact implicit
method for parabolic equations, Journal of Computational Physics 28 (1978) 135-166.

\bibitem{gup_man_ste_84}
M.M. Gupta, R.P. Manohar, J.W. Stephenson, A single cell high order scheme for the convection-diffusion equation with variable coefficients, International Journal for Numerical Methods in Fluids 4 (1984) 641-651.

\bibitem{noy_tan_88}
B.J. Noye, H.H. Tan, A third-order semi-implicit finite difference method
for solving the one-dimensional convection–diffusion equation, International
Journal for Numerical Methods in Engineering 26 (1988) 1615-1629.

\bibitem{den_hud_89}
S.C.R. Dennis, J.D. Hudson, Compact finite difference approximation to operators of Navier–Stokes type, Journal of Computational Physics 85 (1989) 390-416.

\bibitem{spo_car_95}
W.F. Spotz, G.F. Carey, High-order compact scheme for the steady streamfunction
vorticity equations, International Journal for Numerical Methods in
Engineering 38 (1995) 3497-3512.

\bibitem{spo_car_01}
W.F. Spotz, G.F. Carey, Extension of high-order compact schemes to time-dependent problems, Numerical Methods for Partial
Differential Equations 17 (2001) 657-672.

\bibitem{li_tan_for_95}
M. Li, T. Tang, B. Fornberg, A compact fourth-order finite difference scheme for
the incompressible Navier–Stokes equations, International Journal for Numerical
Methods in Fluids 20 (1995) 1137-1151.

\bibitem{zha_98}
J. Zhang, An explicit fourth-order compact finite difference scheme for three dimensional convection–diffusion equation, Communications in Numerical Methods in Engineering 14 (1998) 263-280.

\bibitem{kal_dal_das_02}
J.C. Kalita, D.C. Dalal, A.K. Dass, A class of higher order compact schemes for the
unsteady two-dimensional convection–diffusion equation with variable convection
coefficients, International Journal for Numerical Methods in Fluids 38
(2002) 1111-1131.

\bibitem{deh_04}
M. Dehghan, Numerical solution of the three-dimensional convection–diffusion
equation, Applied Mathematics and Computation 150 (2004) 5-19.

\bibitem{spo_car_98}
W.F. Spotz, G.F. Carey, Formulation and experiments with high-order compact schemes for nonuniform grids, International Journal of Numerical Methods for Heat and Fluid Flow 8
(1998) 288-297.

\bibitem{kal_das_dal_04}
J.C. Kalita, A.K. Dass, D.C. Dalal, A transformation-free HOC scheme for steady convection–diffusion on non-uniform grids, International Journal for Numerical Methods in Fluids 44 (2004) 33-53.

\bibitem{wan_zho_zha_06}
J. Wang, W. Zhong, J. Zhang, High order compact computation and nonuniform grids for stream function vorticity equations, Applied Mathematics and Computation 179
(2006) 108-120.

\bibitem{kar_zha_04}
S. Karaa, J. Zhang, High order ADI method for solving unsteady convection–
diffusion problems, Journal of Computational Physics 198 (2004) 1-9.

\bibitem{you_06}
D. You, A high-order Pad\'{e} ADI method for unsteady convection–diffusion equations,
Journal of Computational Physics 214 (2006) 1-11.

\bibitem{tia_ge_07}
Z.F. Tian, Y.B. Ge, A fourth-order compact ADI method for solving twodimensional
unsteady convection–diffusion problems, Journal of Computational
and Applied Mathematics 198 (2007) 268-286.

\bibitem{tia_11}
Z.F. Tian, A rational high-order compact ADI method for unsteady convection-diffusion equations, Computer Physics Communications 182 (2011) 649-662.

\bibitem{gup_kou_zha_97}
M.M. Gupta, J. Kouatchou, J. Zhang, A compact multigrid solver for convection–diffusion equations, Journal of Computational Physics 132 (1997) 123-129.

\bibitem{gup_zha_00}
M.M. Gupta, J. Zhang, High accuracy multigrid solution of the 3D convection–diffusion equation, Applied Mathematics and Computation 113 (2000) 249-274.

\bibitem{ge_cao_11}
Y.B. Ge, F. Cao, Multigrid method based on the transformation-free HOC scheme on nonuniform grids for 2D convection diffusion problems, Journal of Computational Physics 230 (2011) 4051-4070.

\bibitem{gup_man_79}
M. M. Gupta, R. P. Manohar, Direct solution of the biharmonic equation
using noncoupled approach, Journal of Computational Physics 33 (1979) 236-248.

\bibitem{ste_84}
J. W. Stephenson, Single cell discretization of order two and four for
biharmonic problems, Journal of Computational Physics 55 (1984) 65-80.

\bibitem{book_kel_95}
C.T. Kelly, {\em Iterative Methods for Linear and Nonlinear Equations}, SIAM Publications, Philadelphia, 1995.

\bibitem{ben_cro_fis_05}
M. Ben-Artzi, J-P Croisille, D. Fishelov and S. Trachtenberg,
A pure-compact scheme for the streamfunction formulation of
Navier-Stokes equations, Journal of Computational Physics 205 (2005) 640-664.

\bibitem{kal_gup_10}
J.C. Kalita and M.M. Gupta,
A streamfunction-velocity approach for the 2D transient
incompressible viscous flows, International Journal for Numerical Methods in Fluids 62 (2010) 237-266.

\bibitem{bel_col_89}
J.B. Bell and P. Colella, A second-order projection method for the incompressible Navier-Stokes equations, Journal
of Computational Physics 85 (1989) 257-283.

\bibitem{tia_lia_yu_11}
Z.F. Tian, X. Liang and P. Yu, A higher order compact finite difference algorithm for solving
the incompressible Navier–Stokes equations, International Journal for Numerical Methods in Engineering 88 (2011) 511-532.

\bibitem{ghi_ghi_shi_82}
U. Ghia U, K.N. Ghia, C.T. Shin, High-Re solutions for incompressible flow using the Navier-Stokes equations and
a multigrid method, Journal of Computational Physics 48 (1982) 387-411.

\bibitem{bru_jou_90}
C.H. Bruneau, C. Jouron, An efficient scheme for solving steady incompressible Navier–Stokes equations, Journal
of Computational Physics 89 (1990) 389-413.

\bibitem{gup_kal_05}
M.M. Gupta, J.C. Kalita, A new paradigm for solving Navier-Stokes equations: streamfunction-velocity formulation,
Journal of Computational Physicis 207 (2005) 52-68.

\bibitem{book_roa_98}
P. Roache, {\em Verification and Validation in Computational Science and Egineering},
Hermosa Publishers, Albuquerque, 1998.
\end{thebibliography}

\clearpage
\begin{table}[!h]
\begin{center}
\caption{\sl {Problem 1: $L_1$-, $L_2$-, $L_{\infty}$- norm error and spatial order of convergence with $\delta t=h^2=k^2$}}
{\begin{tabular}{ccccccc} \hline \hline
 Time  &      &$11\times11$ &Order &$21\times21$  &Order   &$41\times41$\\
\hline
$t=0.25$  &$L_1$  &3.635$\times10^{-5}$&3.81  &2.598$\times10^{-6}$&3.91  &1.730$\times10^{-7}$  \\
          &$L_2$  &4.690$\times10^{-5}$&3.84  &3.274$\times10^{-6}$&3.92  &2.156$\times10^{-7}$  \\
   &$L_{\infty}$  &9.758$\times10^{-5}$&3.87  &6.676$\times10^{-6}$&3.94  &4.354$\times10^{-7}$  \\
       &       &                    &      &                    &      &                    \\
$t=0.5$  &$L_1$  &4.277$\times10^{-7}$&3.66  &3.384$\times10^{-8}$&3.84  &2.367$\times10^{-9}$ \\
         &$L_2$  &5.520$\times10^{-7}$&3.69  &4.265$\times10^{-8}$&3.85  &2.951$\times10^{-9}$ \\
  &$L_{\infty}$  &1.150$\times10^{-6}$&3.72  &8.702$\times10^{-8}$&3.87  &5.960$\times10^{-9}$ \\
\hline \hline
\end{tabular}}
\label{table:P1_order_spatial}
\end{center}
\end{table}

\clearpage
\begin{table}[!h]
\begin{center}
\caption{\sl {Problem 1: $L_1$-, $L_2$-, $L_{\infty}$- norm error and temporal order of convergence with $h=k=0.05$}}
{\begin{tabular}{ccccccc} \hline \hline
 Time  &      &$\delta t=0.01$ &Order &$\delta t=0.005$  &Order   &$\delta t=0.0025$\\
\hline
$t=0.25$  &$L_1$  &4.119$\times10^{-5}$&1.98  &1.033$\times10^{-5}$&1.99  &2.598$\times10^{-6}$  \\
          &$L_2$  &5.189$\times10^{-5}$&1.99  &1.302$\times10^{-5}$&1.99  &3.274$\times10^{-6}$  \\
   &$L_{\infty}$  &1.058$\times10^{-4}$&2.00  &2.654$\times10^{-5}$&1.99  &6.676$\times10^{-6}$  \\
       &       &                    &      &                    &      &                    \\
$t=0.5$  &$L_1$  &5.323$\times10^{-7}$&1.99  &1.343$\times10^{-7}$&1.99  &3.384$\times10^{-8}$ \\
         &$L_2$  &6.708$\times10^{-7}$&1.99  &1.693$\times10^{-7}$&1.99  &4.265$\times10^{-8}$ \\
  &$L_{\infty}$  &1.369$\times10^{-6}$&1.99  &3.454$\times10^{-7}$&1.99  &8.702$\times10^{-8}$ \\
\hline \hline
\end{tabular}}
\label{table:P1_order_temporal}
\end{center}
\end{table}

\clearpage
\begin{table}[!h]
\begin{center}
\caption{\sl {Problem 2: $L_1$-, $L_2$-, $L_{\infty}$- norm error and spatial order of convergence. Here $a=100$, $c=d=80$ with $\delta t=h^2=k^2$.}}
{\begin{tabular}{ccccccc} \hline \hline
 Time  &      &$21\times21$ &Order &$41\times41$  &Order   &$81\times81$\\
\hline
$t=0.5$  &$L_1$  &9.902$\times10^{-4}$&4.76  &3.656$\times10^{-5}$&4.10  &2.135$\times10^{-6}$  \\
         &$L_2$  &3.052$\times10^{-3}$&4.24  &1.617$\times10^{-4}$&4.11  &9.394$\times10^{-6}$  \\
   &$L_{\infty}$  &2.836$\times10^{-2}$&3.89  &1.915$\times10^{-3}$&4.11  &1.107$\times10^{-4}$  \\
       &       &                    &      &                    &      &                    \\
$t=1.0$  &$L_1$  &5.363$\times10^{-4}$&4.75  &1.992$\times10^{-5}$&4.05  &1.201$\times10^{-6}$ \\
         &$L_2$  &1.297$\times10^{-3}$&4.23  &6.904$\times10^{-5}$&4.06  &4.134$\times10^{-6}$ \\
  &$L_{\infty}$  &1.036$\times10^{-2}$&4.06  &6.200$\times10^{-4}$&4.00  &3.865$\times10^{-5}$ \\
\hline \hline
\end{tabular}}
\label{table:P2_order_spatial}
\end{center}
\end{table}

\clearpage
\begin{table}[!h]
\begin{center}
\caption{\sl {Problem 2: $L_1$-, $L_2$-, $L_{\infty}$- norm error and temporal order of convergence. Here $a=100$, $c=d=80$ with $h=k=0.05$.}}
{\begin{tabular}{ccccccc} \hline \hline
 Time  &      &$\delta t=0.1$ &Order &$\delta t=0.05$  &Order   &$\delta t=0.025$\\
\hline
$t=0.5$  &$L_1$   &3.608$\times10^{-3}$&1.84  &1.006$\times10^{-3}$&1.96  &2.592$\times10^{-4}$  \\
          &$L_2$  &1.507$\times10^{-2}$&1.79  &4.361$\times10^{-3}$&1.93  &1.147$\times10^{-3}$  \\
   &$L_{\infty}$  &1.551$\times10^{-1}$&2.00  &4.599$\times10^{-2}$&1.90  &1.233$\times10^{-2}$  \\
       &       &                    &      &                    &      &                    \\
$t=1.0$  &$L_1$  &3.736$\times10^{-3}$&2.00  &9.291$\times10^{-4}$&1.98  &2.354$\times10^{-4}$ \\
         &$L_2$  &1.126$\times10^{-2}$&1.86  &3.102$\times10^{-3}$&1.97  &7.928$\times10^{-4}$ \\
  &$L_{\infty}$  &8.383$\times10^{-2}$&1.75  &2.494$\times10^{-2}$&2.02  &6.148$\times10^{-3}$ \\
\hline \hline
\end{tabular}}
\label{table:P2_order_temporal}
\end{center}
\end{table}

\clearpage
\begin{table}[!h]
\begin{center}
\caption{\sl {Problem 3: $L_1$-, $L_2$-, $L_{\infty}$- norm error in $\psi$ and spatial order of convergence with $\delta t=h^2=k^2$}}
{\begin{tabular}{ccccccc} \hline \hline
 Time  &      &$11\times11$ &Order &$21\times21$  &Order   &$41\times41$\\
\hline
$t=0.5$  &$L_1$   &6.403$\times10^{-8}$&3.80  &4.610$\times10^{-9}$&3.90  &3.082$\times10^{-10}$  \\
          &$L_2$  &8.213$\times10^{-8}$&3.83  &5.769$\times10^{-9}$&3.92  &3.814$\times10^{-10}$  \\
   &$L_{\infty}$  &1.661$\times10^{-7}$&3.85  &1.154$\times10^{-8}$&3.93  &7.565$\times10^{-10}$  \\
       &       &                    &      &                    &      &                    \\
$t=1.0$  &$L_1$  &3.936$\times10^{-8}$&3.79  &2.838$\times10^{-9}$&3.90  &1.899$\times10^{-10}$ \\
         &$L_2$  &5.048$\times10^{-8}$&3.83  &3.551$\times10^{-9}$&3.92  &2.350$\times10^{-10}$ \\
  &$L_{\infty}$  &1.019$\times10^{-7}$&3.84  &7.102$\times10^{-9}$&3.93  &4.660$\times10^{-10}$ \\
       &       &                    &      &                    &      &                    \\
$t=1.5$  &$L_1$  &2.400$\times10^{-8}$&3.79  &1.731$\times10^{-9}$&3.90  &1.158$\times10^{-10}$ \\
         &$L_2$  &3.078$\times10^{-8}$&3.83  &2.166$\times10^{-9}$&3.92  &1.434$\times10^{-10}$ \\
  &$L_{\infty}$  &6.212$\times10^{-8}$&3.84  &4.334$\times10^{-9}$&3.93  &2.842$\times10^{-10}$ \\
\hline \hline
\end{tabular}}
\label{table:P3_order_spatial_st}
\end{center}
\end{table}

\clearpage
\begin{table}[!h]
\begin{center}
\caption{\sl {Problem 3: $L_1$-, $L_2$-, $L_{\infty}$- norm error in $\omega$ and spatial order of convergence with $\delta t=h^2=k^2$}}
{\begin{tabular}{ccccccc} \hline \hline
 Time  &      &$11\times11$ &Order &$21\times21$  &Order   &$41\times41$\\
\hline
$t=0.5$   &$L_1$  &1.450$\times10^{-6}$&3.85  &1.004$\times10^{-7}$&3.93  &6.582$\times10^{-9}$  \\
          &$L_2$  &1.819$\times10^{-6}$&3.89  &1.226$\times10^{-7}$&3.92  &7.946$\times10^{-9}$  \\
   &$L_{\infty}$  &3.480$\times10^{-6}$&3.87  &2.340$\times10^{-7}$&3.96  &1.501$\times10^{-8}$  \\
       &       &                    &      &                    &      &                    \\
$t=1.0$  &$L_1$  &8.900$\times10^{-7}$&3.85  &6.172$\times10^{-8}$&3.93  &4.049$\times10^{-9}$ \\
         &$L_2$  &1.116$\times10^{-6}$&3.89  &7.530$\times10^{-8}$&3.95  &4.882$\times10^{-9}$ \\
  &$L_{\infty}$  &2.153$\times10^{-6}$&3.91  &1.428$\times10^{-7}$&3.96  &9.187$\times10^{-9}$ \\
       &       &                    &      &                    &      &                    \\
$t=1.5$  &$L_1$  &5.423$\times10^{-7}$&3.85  &3.763$\times10^{-8}$&3.93  &2.469$\times10^{-9}$ \\
         &$L_2$  &6.795$\times10^{-7}$&3.89  &4.589$\times10^{-8}$&3.95  &2.976$\times10^{-9}$ \\
  &$L_{\infty}$  &1.316$\times10^{-6}$&3.92  &8.692$\times10^{-8}$&3.96  &5.593$\times10^{-9}$ \\
\hline \hline
\end{tabular}}
\label{table:P3_order_spatial_om}
\end{center}
\end{table}

\clearpage
\begin{table}[!h]
\begin{center}
\caption{\sl {Problem 4: $L_1$- and $L_2$- norm difference in horizontal velocity and perceived order of convergence}}
{
\begin{tabular}{cccccc}
\hline
\hline
Time    &Grid Spacing          & $L_1$-norm             &$p$    & $L_2$-norm            &$p$\\
\hline
        &$h=k=\frac{2\pi}{64}$ &                        &       &                       &\\
        &$\delta t=0.005$      &$\|U_3-U_1\|$           &       &$\|U_3-U_1\|$          &\\
        &                      &=6.946$\times10^{-6}$   &       &=1.854$\times10^{-5}$  &\\
$t=1.0$ &$h=k=\frac{2\pi}{128}$&                        &3.19   &                       &4.00\\
        &$\delta t=0.0025$     &                        &       &                       &\\
        &                      &$\|U_3-U_2\|$           &       &$\|U_3-U_2\|$          &\\
        &$h=k=\frac{2\pi}{256}$&=6.878$\times10^{-7}$   &       &=1.092$\times10^{-6}$  &\\
        &$\delta t=0.0005$     &                        &       &                       &\\
\hline
        &$h=k=\frac{2\pi}{64}$ &                        &       &                       &\\
        &$\delta t=0.005$      &$\|U_3-U_1\|$           &       &$\|U_3-U_1\|$          &\\
        &                      &=1.766$\times10^{-3}$   &       &=3.203$\times10^{-3}$  &\\
$t=5.0$ &$h=k=\frac{2\pi}{128}$&                        &3.59   &                       &3.28\\
        &$\delta t=0.0025$     &                        &       &                       &\\
        &                      &$\|U_3-U_2\|$           &       &$\|U_3-U_2\|$          &\\
        &$h=k=\frac{2\pi}{256}$&=1.354$\times10^{-4}$   &       &=2.992$\times10^{-4}$  &\\
        &$\delta t=0.0005$     &                        &       &                       &\\
\hline
        &$h=k=\frac{2\pi}{64}$ &                        &       &                       &\\
        &$\delta t=0.005$      &$\|U_3-U_1\|$           &       &$\|U_3-U_1\|$          &\\
        &                      &=1.155$\times10^{-2}$   &       &=1.627$\times10^{-2}$  &\\
$t=10.0$&$h=k=\frac{2\pi}{128}$&                        &1.86   &                       &1.64\\
        &$\delta t=0.0025$     &                        &       &                       &\\
        &                      &$\|U_3-U_2\|$           &       &$\|U_3-U_2\|$          &\\
        &$h=k=\frac{2\pi}{256}$&=2.498$\times10^{-3}$   &       &=3.962$\times10^{-3}$  &\\
        &$\delta t=0.0005$     &                        &       &                       &\\
\hline
\end{tabular}}
\label{table:P4_u_order}
\end{center}
\end{table}

\clearpage
\begin{table}[!h]
\begin{center}
\caption{\sl {Problem 4: $L_1$- and $L_2$- norm difference in vertical velocity and perceived order of convergence}}
{
\begin{tabular}{cccccc}
\hline
\hline
Time    &Grid Spacing          & $L_1$-norm             &$p$    & $L_2$-norm            &$p$\\
\hline
        &$h=k=\frac{2\pi}{64}$ &                        &       &                       &\\
        &$\delta t=0.005$      &$\|V_3-V_1\|$           &       &$\|V_3-V_1\|$          &\\
        &                      &=2.041$\times10^{-4}$   &       &=5.063$\times10^{-4}$  &\\
$t=1.0$ &$h=k=\frac{2\pi}{128}$&                        &4.10   &                       &4.12\\
        &$\delta t=0.0025$     &                        &       &                       &\\
        &                      &$\|V_3-V_2\|$           &       &$\|V_3-V_2\|$          &\\
        &$h=k=\frac{2\pi}{256}$&=1.128$\times10^{-5}$   &       &=2.750$\times10^{-5}$  &\\
        &$\delta t=0.0005$     &                        &       &                       &\\
\hline
        &$h=k=\frac{2\pi}{64}$ &                        &       &                       &\\
        &$\delta t=0.005$      &$\|V_3-V_1\|$           &       &$\|V_3-V_1\|$          &\\
        &                      &=2.872$\times10^{-3}$   &       &=5.589$\times10^{-3}$  &\\
$t=5.0$ &$h=k=\frac{2\pi}{128}$&                        &3.05   &                       &2.51\\
        &$\delta t=0.0025$     &                        &       &                       &\\
        &                      &$\|V_3-V_2\|$           &       &$\|V_3-V_2\|$          &\\
        &$h=k=\frac{2\pi}{256}$&=3.094$\times10^{-4}$   &       &=8.369$\times10^{-4}$  &\\
        &$\delta t=0.0005$     &                        &       &                       &\\
\hline
        &$h=k=\frac{2\pi}{64}$ &                        &       &                       &\\
        &$\delta t=0.005$      &$\|V_3-V_1\|$           &       &$\|V_3-V_1\|$          &\\
        &                      &=1.234$\times10^{-2}$   &       &=1.733$\times10^{-2}$  &\\
$t=10.0$&$h=k=\frac{2\pi}{128}$&                        &2.12   &                       &1.99\\
        &$\delta t=0.0025$     &                        &       &                       &\\
        &                      &$\|V_3-V_2\|$           &       &$\|V_3-V_2\|$          &\\
        &$h=k=\frac{2\pi}{256}$&=2.305$\times10^{-3}$   &       &=3.494$\times10^{-3}$  &\\
        &$\delta t=0.0005$     &                        &       &                       &\\
\hline
\end{tabular}}
\label{table:P4_v_order}
\end{center}
\end{table}

\clearpage
\begin{table}
\begin{center}
\caption{Problem 5: Properties of Primary, Secondary and Tertiary Vortices for
the lid-driven square cavity from $Re=1000$ to $Re=7500$.}
\begin{footnotesize}
{\begin{tabular}{ccccccccc}\hline
Vortex &Property        &Ref.                   &$Re=$1000      &$Re=$3200      &$Re=$5000  &$Re=$7500\\
\hline
Primary &$\psi_{min}$   &                       &-0.1180        &-0.1214        &-0.1218        &-0.1220\\
        &               &\cite{ghi_ghi_shi_82}  &-0.1179        &-0.1203        &-0.1190        &-0.1120\\
        &               &\cite{bru_jou_90}      &-0.1163        &---            &-0.1142        &-0.1113\\
        &               &\cite{gup_kal_05}      &-0.117         &-0.122         &-0.122         &-0.122\\
        &($x,y$)        &                       &0.5313, 0.5625 &0.5156, 0.5391 &0.5156, 0.5312 &0.5156, 0.5312\\
        &               &\cite{ghi_ghi_shi_82}  &0.5313, 0.5625 &0.5165, 0.5469 &0.5117, 0.5352 &0.5117, 0.5322 \\
        &               &\cite{bru_jou_90}      &0.5313, 0.5586 &---            &0.5156, 0.5313 &0.5156, 0.5234\\
        &               &\cite{gup_kal_05}      &0.5250, 0.5625 &0.5188, 0.5438 &0.5125, 0.5375 &0.5125, 0.5313\\
Secondary &             &                       &               &               &               &\\
TL &$\psi_{max}$        &                       &               &8.10e-4        &1.60e-3        &2.36e-3 \\
        &               &\cite{ghi_ghi_shi_82}  &               &7.28e-4        &1.46e-3        &2.05e-3\\
        &               &\cite{bru_jou_90}      &               &---            &1.75e-3        &3.14e-3\\
        &               &\cite{gup_kal_05}      &               &7.33e-4        &1.54e-3        &2.07e-3\\
        &($x,y$)        &                       &               &0.0549, 0.8984 &0.0625, 0.9062 &0.0703, 0.9063\\
        &               &\cite{ghi_ghi_shi_82}  &               &0.0547, 0.8984 &0.0625, 0.9102 &0.0664, 0.9141\\
        &               &\cite{bru_jou_90}      &               &---            &0.0625, 0.9102 &0.0664, 0.9141\\
        &               &\cite{gup_kal_05}      &               &0.0563, 0.9000 &0.0688, 0.9125 &0.0688, 0.9125\\
BL      &$\psi_{max}$   &                       &2.53e-4        &1.13e-3        &1.42e-3        &1.63e-3\\
        &               &\cite{ghi_ghi_shi_82}  &2.31e-4        &9.78e-4        &1.36e-3        &1.47e-3\\
        &               &\cite{bru_jou_90}      &3.25e-4        &---            &2.22e-3        &1.76e-3\\
        &               &\cite{gup_kal_05}      &2.02e-4        &1.03e-3        &1.32e-3        &1.60e-3\\
        &($x,y$)        &                       &0.0782, 0.0781 &0.0804, 0.1203 &0.0703, 0.1406 &0.0625, 0.1563\\
        &               &\cite{ghi_ghi_shi_82}  &0.0859, 0.0781 &0.0859, 0.1094 &0.0703, 0.1367 &0.0645, 0.1504  \\
        &               &\cite{bru_jou_90}      &0.0859, 0.0820 &---            &0.0664, 0.1484 &0.0703, 0.1289\\
        &               &\cite{gup_kal_05}      &0.0875, 0.0750 &0.0813, 0.1188 &0.0750, 0.1313 &0.0688, 0.1500\\
BR      &$\psi_{max}$   &                       &1.88e-3        &2.93e-3        &3.25e-3        &3.49e-3\\
        &               &\cite{ghi_ghi_shi_82}  &1.75e-3        &3.14e-3        &3.08e-3        &3.28e-3\\
        &               &\cite{bru_jou_90}      &1.91e-3        &---            &4.65e-3        &8.32e-3\\
        &               &\cite{gup_kal_05}      &1.70e-3        &2.86e-3        &2.96e-3        &3.05e-3\\
        &($x,y$)        &                       &0.8594, 0.1094 &0.8203, 0.0859 &0.7969, 0.0703 &0.7813, 0.0625 \\
        &               &\cite{ghi_ghi_shi_82}  &0.8594, 0.1094 &0.8125, 0.0859 &0.8056, 0.0742 &0.7813, 0.0625 \\
        &               &\cite{bru_jou_90}      &0.8711, 0.1094 &---            &0.8301, 0.0703 &0.8828, 0.0820\\
        &               &\cite{gup_kal_05}      &0.8625, 0.1125 &0.8125, 0.0875 &0.8000, 0.0750 &0.7813, 0.0625\\
Tertiary&               &                       &               &               &               &\\
BR      &$\psi_{min}$   &                       &               &-3.90e-7       &-3.04e-6       &-6.35e-5\\
        &               &\cite{ghi_ghi_shi_82}  &               &-2.52e-7       &-1.43e-6       &-3.28e-5\\
        &               &\cite{bru_jou_90}      &               &---            &-2.47e-5       &---\\
        &               &\cite{gup_kal_05}      &               &-2.37e-7       &-1.70e-6       &-1.89e-5\\
        &($x,y$)        &                       &               &0.9880, 0.0114 &0.9744, 0.0211 &0.9453, 0.0469 \\
        &               &\cite{ghi_ghi_shi_82}  &               &0.9844, 0.0078 &0.9805, 0.0195 &0.9492, 0.0430 \\
        &               &\cite{bru_jou_90}      &               &---            &0.9668, 0.0293 &---\\
        &               &\cite{gup_kal_05}      &               &0.9875, 0.0125 &0.9750, 0.0188 &0.9500, 0.0375\\
\hline
Grid Size&              &                       &$65 \times 65$  &$129 \times 129$&$129 \times 129$&$129 \times 129$ \\
        &               &\cite{ghi_ghi_shi_82}  &$129 \times 129$&$129 \times 129$&$257 \times 257$&$257 \times 257$\\
        &               &\cite{bru_jou_90}      &$257 \times 257$&$257 \times 257$&$257 \times 257$&$257 \times 257$\\
        &               &\cite{gup_kal_05}      &$81\times81$    &$161\times161$  &$161\times161$  &$161\times161$\\
\hline
\end{tabular}}
\label{table:P5_com}
\end{footnotesize}
\end{center}
\end{table}

\clearpage
\clearpage

\clearpage

\clearpage

\clearpage

\clearpage

\clearpage

\clearpage

\clearpage

\clearpage

\end{document}